\documentclass[namedreferences]{solarphysics}
\usepackage[optionalrh,solaenum]{spr-sola-addons} 
\usepackage{graphicx}                    
\usepackage{amssymb}                    
\usepackage{color}                       
\usepackage{url}                         
\usepackage{lscape}
\usepackage{hangcaption}
\usepackage{longtable}

\begin{document}

\begin{article}

\begin{opening}

\title{Radial Speed Evolution of Interplanetary Coronal Mass Ejections During Solar Cycle 23}

%
\author{T.~\surname{Iju}$^{1}$\sep
        M.~\surname{Tokumaru}$^{1}$\sep
        K.~\surname{Fujiki}$^{1}$      
       }

%

%
  \institute{$^{1}$ Solar-Terrestrial Environment Laboratory, Nagoya University, Furo-cho, Chikusa-ku, Nagoya, Aichi 464-8601, Japan \\
  email: \url{tomoya@stelab.nagoya-u.ac.jp} \\ 
             }

\begin{abstract}
We report radial-speed evolution of interplanetary coronal mass ejections 
(ICMEs) detected by the \textit{Large Angle and Spectrometric Coronagraph} 
onboard the \textit{Solar and heliospheric Observatory} (SOHO/LASCO), 
interplanetary scintillation (IPS) at 327 MHz,and \textit{in-situ} observations. 
We analyzed solar-wind disturbance factor (\textit{g}-value) data derived from IPS 
observations during 1997\,--\,2009 covering nearly whole period of Solar Cycle 23. 
By comparing observations from SOHO/LASCO, IPS, and \textit{in situ}, 
we identified 39 ICMEs that could be analyzed carefully. Here, we defined two speeds 
[${V_{\mathrm{SOHO}}}$ and ${V_{\mathrm{bg}}}$], which are initial speed of 
the ICME and the speed of the background solar wind, respectively. 
Examination of these speeds yield the following results: 
i) Fast ICMEs (with $V_{\mathrm{SOHO}} - V_\mathrm{bg} > 500$ ${\mathrm{km~s^{-1}}}$) 
rapidly decelerate, moderate ICMEs 
(with $0$ $\mathrm{km~s^{-1}}$ $\le V_{\mathrm{SOHO}} - V_\mathrm{bg} \le 500$ ${\mathrm{km~s^{-1}}}$) 
show either gradually decelerating or uniform motion, and slow ICMEs 
(with $V_{\mathrm{SOHO}} - V_\mathrm{bg} <$ $0$ ${\mathrm{km~s^{-1}}}$) accelerate. 
The radial speeds converge on the speed of the background solar wind during their outward propagation. 
We subsequently find; ii) both the acceleration and deceleration are nearly 
complete by $0.79 \pm 0.04$ AU, and those are ended when the ICMEs reach 
a $489 \pm 21$ $\mathrm{km~s^{-1}}$. 
iii) For ICMEs with $V_{\mathrm{SOHO}} - V_\mathrm{bg} \ge$ $0$ ${\mathrm{km~s^{-1}}}$, 
\textit{i.e.} fast and moderate ICMEs, a linear equation $a = -{\gamma}_{\mathrm{1}}(V - V_\mathrm{bg})$ 
with ${\gamma}_{\mathrm{1}} = 6.58 \pm 0.23 \times 10^{-6}$ ${\mathrm{s^{-1}}}$ is more appropriate 
than a quadratic equation $a = -{\gamma}_{\mathrm{2}}(V - V_\mathrm{bg})|V - V_\mathrm{bg}|$ 
to describe their kinematics, where ${{\gamma}_\mathrm{1}}$ and ${{\gamma}_\mathrm{2}}$ are 
coefficients, and $a$ and $V$ are the acceleration and ICME speed, respectively, because the ${\chi^{\mathrm{2}}}$ for 
the linear equation satisfies the statistical significance level of 0.05, while the quadratic one does not. 
These results support the assumption that the radial motion of ICMEs is governed by a drag force 
due to interaction with the background solar wind. These findings also suggest 
that ICMEs propagating faster than the background solar wind are controlled mainly by 
the hydrodynamic Stokes drag. 
\end{abstract}

%

\end{opening}

~\\
\textbf{Abbreviations}\\
ACE~~~~~~~~\textit{Advanced Composition Explorer}\\
AU~~~~~~~~~~Astronomical unit\\
CC~~~~~~~~~~Correlation coefficient\\
CDAW~~~~~Coordinated Data Analysis Workshop\\
CME~~~~~~~~Coronal mass ejection\\
CPI~~~~~~~~~\textit{Comprehensive Plasma Instrumentation}\\
ESA~~~~~~~~~European Space Agency\\
FOV~~~~~~~~~Field-of-view\\
GSFC~~~~~~~Goddard Space Flight Center\\
ICME~~~~~~~Interplanetary coronal mass ejection\\
IDED~~~~~~~IPS disturbance event day\\
IDEDs~~~~~~IPS disturbance event days\\
IMP~~~~~~~~~\textit{Interplanetary Monitoring Platform}\\
IPS~~~~~~~~~~Interplanetary Scintillation\\
LASCO~~~~\textit{Large Angle and Spectrometric Coronagraph}\\
LOS~~~~~~~~~Line-of-sight\\
MIT~~~~~~~~\textit{Massachusetts Institute of Technology Faraday Cup Experiment}\\
NASA~~~~~~National Aeronautics and Space Administration\\
OMNI~~~~~~Operating Missions as Nodes on the Internet\\
SOHO~~~~~~\textit{Solar and Heliospheric Observatory}\\
STEL~~~~~~~Solar-Terrestrial Environment Laboratory\\
STEREO~~\textit{Solar-Terrestrial Relations Observatory}\\
SWE~~~~~~~~\textit{Solar Wind Experiment}\\
SWEPAM~\textit{Solar Wind Electron, Proton, and Alpha Monitor}

 \section{Introduction}
         \label{introduction} 

Coronal mass ejections (CMEs) are transient events in which large amounts of plasma are ejected 
from the solar corona (\textit{e.g.} \opencite{Gosling1974}). Interplanetary counterparts 
of CMEs are called interplanetary coronal mass ejections (ICMEs). Since ICMEs 
seriously affect the space environment around the Earth, understanding of their fundamental 
physics, \textit{e.g.} generation, propagation, and interaction with the Earth's magnetosphere, 
is very important for space-weather forecasting (\textit{e.g.} \opencite{Tsurutani1988}; 
\opencite{Gosling1990}). In particular, the dynamics of ICME propagation is one of the key 
pieces of information for predicting geomagnetic storms. 

Propagation of ICMEs has been studied by various methods. Earlier studies combining 
space-borne coronagraphs with \textit{in-situ} observations revealed that ICME speeds significantly 
evolve between near-Sun and 1 AU. \inlinecite{Schwenn1983} reported the correlation between CMEs 
and interplanetary disturbances using the \textit{P78-1/Solwind} coronagraph, the \textit{Helios-1} 
and -2 solar probes, and a ground-based H$\alpha$ coronagraph. He showed that fast CMEs associated 
with flares exhibit no acceleration into interplanetary space, while slow CMEs related to 
prominence eruptions accelerate. \inlinecite{Lindsay1999} examined the relation between propagation 
speeds of CMEs observed by the \textit{Solwind} coronagraph and \textit{Solar Maximum Mission} 
coronagraph/polarimeter and those of the ICMEs observed by the \textit{Helios-1} and 
\textit{Pioneer Venus Orbiter} for 31 CMEs and their associated ICMEs. They found a good correlation between 
the speeds of CMEs and those of ICMEs observed in interplanetary space between 
0.7 and 1 AU. They also found that the speeds of most ICMEs range from 380 ${\mathrm{km~s^{-1}}}$ to 
600 ${\mathrm{km~s^{-1}}}$, while CME speeds show a wider range of from $\approx 10$ ${\mathrm{km~s^{-1}}}$ 
to 1500 ${\mathrm{km~s^{-1}}}$. These findings suggest that the ICME speeds tend to converge to an average solar-wind 
speed as they propagate through interplanetary space. \inlinecite{Gopalswamy2000} determined 
an effective acceleration for 28 CMEs observed by the \textit{Large Angle and Spectrometric Coronagraph} 
(LASCO: \opencite{Brueckner1995}) onboard the \textit{Solar and Heliospheric Observatory} 
(SOHO) spacecraft between 1996 and 1998. On the assumption that the acceleration is constant, 
they found a very good anti-correlation between the accelerations and initial speeds of 
CMEs, and a critical speed of 405 ${\mathrm{km~s^{-1}}}$; this value is close to the typical speed of the solar 
wind in the equatorial plane. Following this research, \inlinecite{Gopalswamy2001} described 
an empirical model for predicting of arrival of the ICMEs at 1 AU; this model is based on 
their previous work \cite{Gopalswamy2000} and its accuracy is improved by allowing for 
cessation of the interplanetary acceleration before 1 AU. They showed that the acceleration 
cessation distance is 0.76 AU, and this result agrees reasonably well with observations by 
SOHO, \textit{Advanced Composition Explorer} (ACE: \opencite{Stone1998}), and other spacecraft at 
1 AU. 

We expect that the acceleration or deceleration of ICMEs is controlled by a drag force 
caused by interaction between ICMEs and the solar wind. \inlinecite{Vrsnak2002} 
proposed an advanced model for the motion of ICMEs; this model considered the interaction 
with solar wind using a simple expression for the acceleration: $a = -{\gamma}_{\mathrm{1}}(V - V_\mathrm{bg})$, 
where ${\gamma}_{\mathrm{1}}$, $V$, and $V_\mathrm{bg}$ are the coefficient, ICME speed, 
and speed of the background solar wind, respectively. They also compared their model 
with a drag-acceleration model $a = -{\gamma}_{\mathrm{2}}(V - V_\mathrm{bg})|V - V_\mathrm{bg}|$, 
where ${\gamma}_{\mathrm{2}}$ is the coefficient for this equation; this expression is 
known as the aerodynamic drag force (\textit{e.g.} \opencite{Chen1996}; \opencite{Cargill2004}). 
Both models have been tested by comparing with CME observations. \inlinecite{Tappin2006} studied 
the propagation of a CME that occurred on 5 April 2003 using observations by the SOHO/LASCO, 
the \textit{Solar Mass Ejection Imager} onboard the \textit{Coriolis} satellite, and the \textit{Ulysses} 
spacecraft. \inlinecite{Maloney2010} derived the three-dimensional kinematics for three ICMEs detected 
between 2008 and 2009 using the \textit{Solar-Terrestrial Relations Observatory-A} (STEREO-A) 
and -B spacecraft observations. \inlinecite{Temmer2011} examined the influence of the solar 
wind on the propagation of some ICMEs using the STEREO-A and -B spacecraft. 
Although the propagation of ICMEs has been studied by many investigators, their dynamics 
is still not well understood. This is mainly due to the lack of observational data 
about ICMEs between 0.1 and 1 AU. Almost all ICME observations are currently limited to 
the near-Earth area in the equatorial plane. 

Remote sensing using radio waves is a suitable method for collecting global data on 
ICMEs. For example, \inlinecite{Reiner2007} derived kinematical parameters 
for 42 ICME/shocks from measurements of type-II radio emission. \inlinecite{Woo1988} studied the 
shock propagation using Doppler-scintillation measurements of radio waves emitted from 
planetary spacecraft, and showed the speed profiles of shocks between 0.05 and 0.93 AU. In 
addition to these measurements, interplanetary scintillation (IPS) is a type of remote sensing. 
IPS is a phenomenon where signals from a point-like radio source, such as quasars and 
active galactic nuclei, fluctuate due to density irregularities in the solar wind \cite{Hewish1964}. 
IPS observations allow us to probe the inner heliosphere using many radio sources, and 
this is a useful means to study the global structure and propagation dynamics 
of ICMEs in the solar wind (\textit{e.g.} \opencite{Gapper1982}; \opencite{Tappin1983}; 
\opencite{Watanabe1986}; \opencite{Tokumaru2000a}, \citeyear{Tokumaru2003}; \opencite{Manoharan2000}; 
\opencite{Jackson2002}; \opencite{Bisi2010}; \opencite{Jackson2011}; \opencite{Manoharan2010}; see also 
\opencite{Watanabe1989}). For the kinematics of interplanetary disturbances, \inlinecite{Vlasov1992} 
reported that the radial dependence of speed can be represented by a power-law function [$V \approx R^{-\alpha}$] 
with $\alpha$ in the range $0.25 < \alpha < 1$ from analysis of all-sky scintillationindices 
maps. \inlinecite{Manoharan2006} examined radial evolution of 30 CMEs observed by 
SOHO/LASCO, ACE, and the Ooty radio-telescope between 1998 and 2004. He showed 
that most CMEs tend to attain the speed of the ambient flow at 1 AU and also reported a 
power-law form of radial-speed evolution for these events. 

We take advantage of IPS observation to determine the ICME speed and acceleration. In 
the current study, we analyze the solar-wind disturbance factor (\textit{g}-value) derived from IPS 
observations during 1997\,--\,2009 covering nearly the whole of Solar Cycle 23 and make a list 
of disturbance event days in the period. We define an ``ICME'' as a series of events including 
a near-Sun CME, an interplanetary disturbance, and a near-Earth ICME in this study. By 
comparing our list with that of CME/ICME pairs, we identify many events that are detected 
at three locations between the Sun and the Earth's orbit, \textit{i.e.} near-Sun, interplanetary space, 
and near-Earth, and derive their radial speed profiles. We then analyze the relationship between 
the acceleration and speed difference for the ICMEs. The outline of this article is as 
follows: Section~\ref{observation} describes the IPS observations made with the 327 MHz radio-telescope 
system of the Solar-Terrestrial Environment Laboratory (STEL), Nagoya University. Section~\ref{method} 
describes the criteria for ICME identification and the method for estimating ICME 
speeds and accelerations between the corona and 1 AU. Section~\ref{results} provides the radial-speed 
profiles of ICMEs and the analyses of the propagation properties. Section~\ref{discussion} discusses the 
results, while Section~\ref{conclusion} summarizes the main conclusions of our study.

\section{STEL IPS Observation} 
      \label{observation}      

STEL IPS observations have been carried out regularly since the early 1980s using multiple 
ground-based radio-telescope stations operated at 327 MHz (\opencite{Kojima1990}; \opencite{Asai1995}). 
The IPS observations at 327 MHz allow us to determine the solar-wind 
condition between 0.2 and 1 AU with a cadence of 24 hours. In our observations, nearly 
30 radio sources within a solar elongation of $60^{\circ}$ are observed daily between April and December. 
The IPS observations on a given day are made when each radio source traverses 
the local meridian.

The solar-wind speed and disturbance factor, the so called ``\textit{g}-value'' \cite{Gapper1982}, 
are derived from IPS observations. A \textit{g}-value is calculated for each source using 
the following equation: 
\begin{equation}
  \label{eq.gvalue}
g = \frac{{\Delta}S}{{\Delta}S_{\mathrm{m}}(\varepsilon)}, 
\end{equation}
where ${{\Delta}S}$ and ${{\Delta}S_{\mathrm{m}}(\varepsilon)}$ are the observed fluctuation level of 
radio signals and their yearly mean, respectively. ${{\Delta}S_{\mathrm{m}}(\varepsilon)}$ varies with 
the solar elongation angle [$\varepsilon$] for a line-of-sight (LOS) from an observed radio source 
to a telescope. When a radio signal is weakly scattered, the \textit{g}-value is given by 
the following equation (\opencite{Tokumaru2003}, \citeyear{Tokumaru2006}): 
\begin{equation} 
  \label{eq.gvalue2} 
g^{2} = \frac{1}{K} \int_{0}^{\infty} \mathrm{d}z{\{}{\Delta}N_\mathrm{e}{\}}^{2}{\omega}(z), 
\end{equation} 
here, $z$ is the distance along a LOS, ${{\Delta}N_\mathrm{e}}$ is the fluctuation level of 
solar-wind (electron) density, $K$ is the normalization factor based on the mean density 
fluctuation of the background solar wind, and ${\omega}(z)$ is the IPS weighting function 
\cite{Young1971}. We note that ${{\Delta}N_\mathrm{e}}$ is nearly proportional to 
the solar-wind density [${N_\mathrm{e}}$]; ${\Delta}N_\mathrm{e} \propto N_\mathrm{e}$ \cite{Coles1978}, 
and the weak-scattering condition holds for $R > 0.2$ AU, where $R$ is 
the radial distance from the Sun.

        %
         \begin{figure}
         \centerline{\includegraphics[width=0.41 \textwidth,clip=]{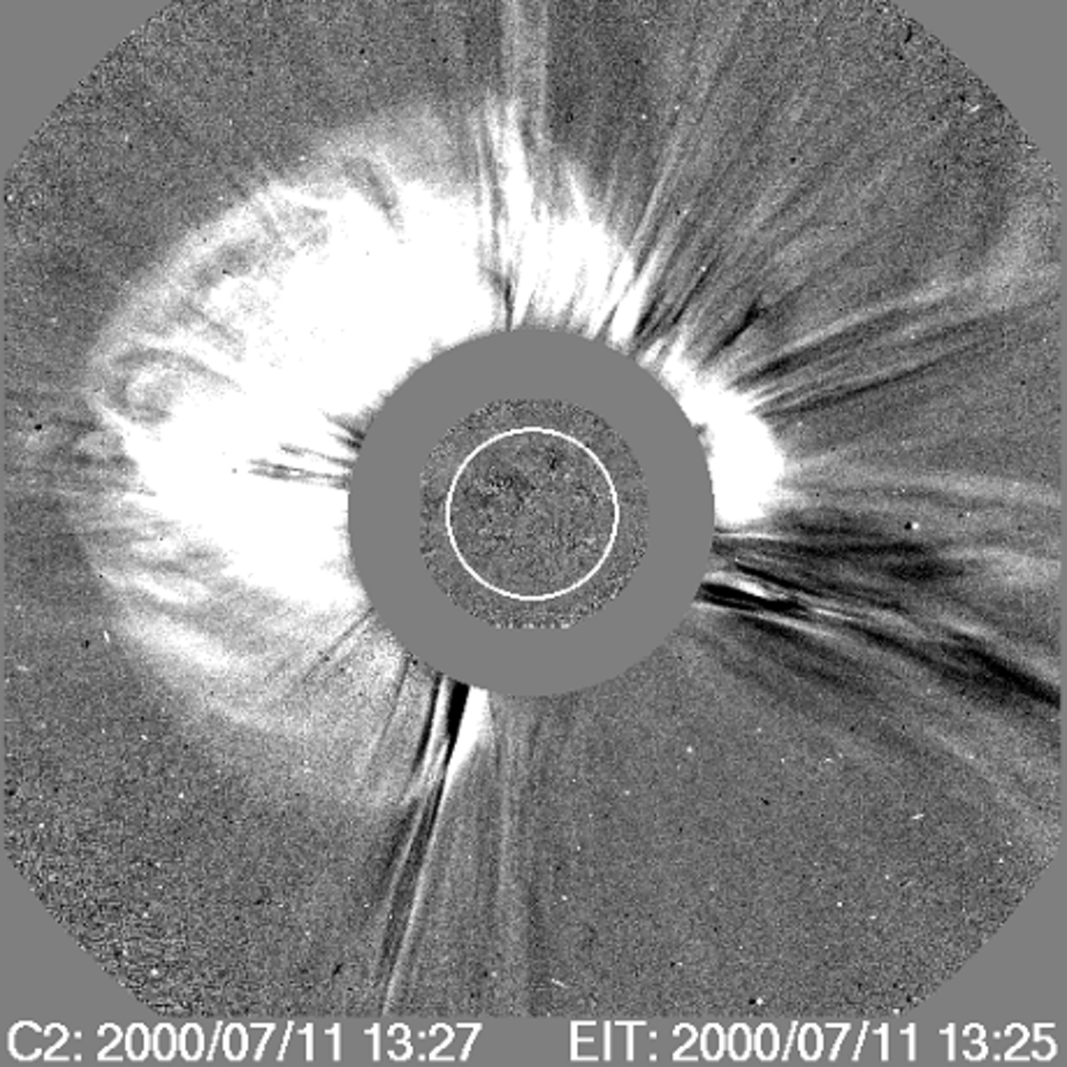}
                     \hspace*{0.03 \textwidth}
                     \includegraphics[width=0.43 \textwidth,clip=]{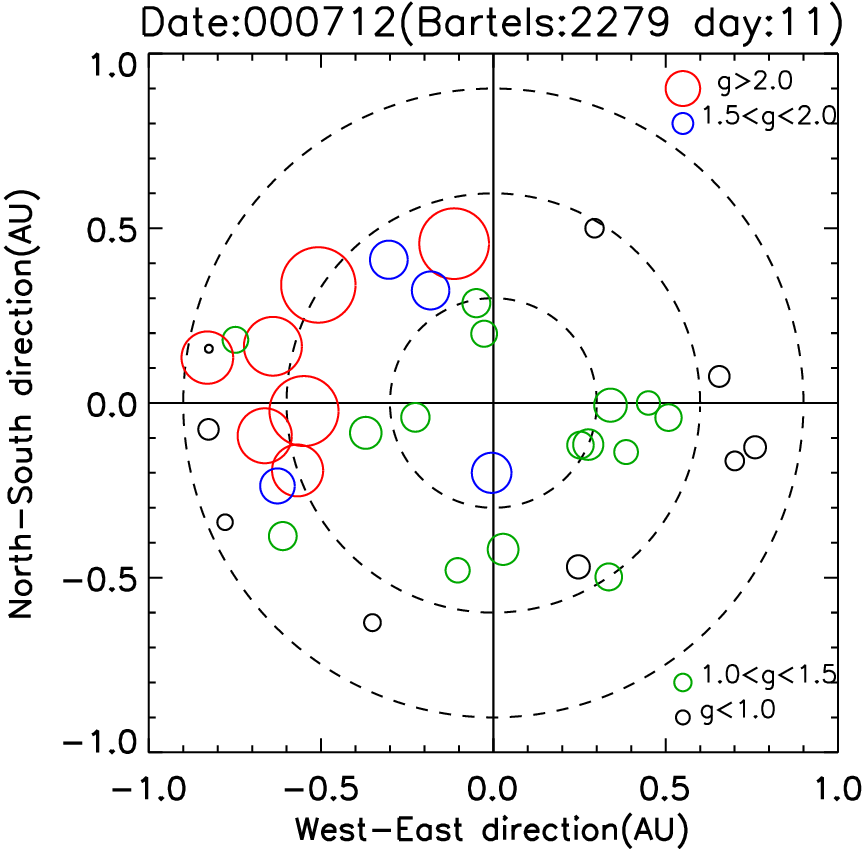}
                     }
           \vspace{-0.05 \textwidth}
           \centerline{\small \bf     
                       \hspace{0.0 \textwidth}  \color{black}{(a)}
                       \hspace{0.42 \textwidth}  \color{black}{(b)}
                       \hfill
                       }
           \vspace{0.015 \textwidth} 
         \caption{
         (a) White-light difference image for the halo CME on 11 July 2000 from the SOHO/LASCO-C2 
         coronagraph (\protect\url{cdaw.gsfc.nasa.gov/CME_list/index.html}), and 
        (b) \textit{g}-map obtained from our IPS observations on 12 July 2000. The \textit{g}-map center 
        corresponds to the location of the Sun, and concentric circles indicate radial distances of 0.3 AU, 0.6 AU, 
        and 0.9 AU. Colored open solid circles indicate the locations of the closest point to the Sun 
        (the P-point) on the LOS for the radio sources in the sky plane. The center of the colored circle 
        indicates the heliocentric distance of the P-point on the LOS, and color and diameter represent 
        the \textit{g}-value level for each source. We use four bins of $g < 1.0$ (black), $1.0 < g < 1.5$ (green), 
        $1.5 < g < 2.0$ (blue), and $g > 2.0$ (red) for the \textit{g}-map. A group of P-points with red or blue circles 
        indicates a disturbance related to the 11 July 2000 CME.
        }
         \label{fig1}
         \end{figure}
        
A \textit{g}-value represents the relative level of density fluctuation integrated along a LOS. 
For quiet solar-wind conditions, the \textit{g}-value is around unity. With dense plasma or high 
turbulence as an ICME passes across a LOS, the \textit{g}-value becomes greater than unity because 
of the ${{\Delta}N_\mathrm{e}}$ ($\propto N_\mathrm{e}$) increase. In contrast, a \textit{g}-value 
less than unity indicates a rarefaction of the solar wind. Hence, detecting an abrupt increase 
in \textit{g}-value is a useful means to detect an ICME. 

The location of the LOS for a radio source exhibiting a \textit{g}-value enhancement in the sky 
plane indicates a turbulent region is present. A sky-map of enhanced \textit{g}-values for the sources 
observed in a day is called a ``\textit{g}-map'' \cite{Gapper1982,Hewish1986}. 
This map provides information on the spatial distribution of ICMEs. Figure 1 shows an 
example of a \textit{g}-map for a CME event. A white-light difference image of a CME observed by 
the SOHO/LASCO-C2 coronagraph is shown in the left-hand panel of Figure~\ref{fig1}. As shown 
here, a bright balloon-like structure was observed on the northeast limb on 11 July 2000. 
This event was reported as an asymmetric halo CME in the SOHO/LASCO CME Catalog 
(\opencite{Yashiro2004}; \opencite{Gopalswamy2009}; available at \url{cdaw.gsfc.nasa.gov/CME_list/}). 
The right-hand panel of Figure~\ref{fig1} is a \textit{g}-map derived from our IPS observation on 12 July 
2000. The center of the map corresponds to the location of the Sun, and the horizontal 
and vertical axes are parallel to the East--West and North--South directions, respectively. 
The concentric circles indicate the radial distances to the closest approach of the LOS of 
0.3 AU, 0.6 AU, and 0.9 AU. The radial distance [$r_\mathrm{IPS}$] for each LOS is given by 
$r_\mathrm{IPS} = r_{\mathrm{E}} \sin{\varepsilon}$, where $r_\mathrm{E}$ is the distance between 
the Sun and the Earth, \textit{i.e.} 1 AU and ${\varepsilon}$ is the solar 
elongation angle for the LOS. This calculation is based on the approximation that a large 
fraction of IPS is given by the wave scattering at the closest point to the Sun (the P-point) 
on a LOS \cite{Hewish1964}. Since ten LOSs between 0.4 and 0.7 AU in 
the eastern hemisphere (left-hand side of \textit{g}-map) exhibit high \textit{g}-values, a group of them is 
considered as the interplanetary counterpart of the 11 July 2000 CME event. This CME was 
also detected by \textit{in-situ} observation at 1 AU on 13 July 2000 and reported as a near-Earth 
ICME \cite{Richardson2010}. 
In this way, a \textit{g}-map can visualize an ICME between 0.2 and 1 AU. The \textit{g}-value data 
have been available from our IPS observation since 1997 \cite{Tokumaru2000b}. To find 
the \textit{g}-value enhancements due to ICMEs from the \textit{g}-value data obtained between 1997 and 
2009, we define criteria for the ICME identifications as mentioned in the next section. 

\section{Method} 
      \label{method}

\subsection{ICME Identification} 
      \label{identification}

First, we define disturbance days due to an ICME in the IPS data. In this determination, we 
consider a threshold \textit{g}-value and the number of sources exhibiting the threshold or beyond. 
The average [$a_{g}$] and standard deviation [${\sigma}_{g}$] for the \textit{g}-values obtained 
by STEL IPS observations between 1997 and 2009 are 1.07 and 0.47, respectively. From these, we regard a 
\textit{g}-value for a disturbed condition on a given day to be $a_{g} + {\sigma}_{g}$ or more, and we decide to 
use 1.5 as this threshold. We also define an ``observation day'' as a day on which 15 or more 
sources are observed by our radio-telescope system; this minimum number is equal to half 
the mean number of sources observed in a day. In an observation day, when five or more 
sources showed a disturbed condition, we judge that a disturbance had occurred. Combining 
the above criteria, we define an ``IPS disturbance event day'' (IDED) as a day on which 
$g \ge 1.5$ sources numbered five or more on an observation day. Using this definition, we find 
656 IDEDs in our period of research. From these, we eliminate periods with four or more 
consecutive IDEDs because they are likely related to co-rotating stream interaction regions 
\cite{Gapper1982}. However, we do not eliminate two periods including the 2000 Bastille 
Day (illustrated in Figure~\ref{fig1}) and 2003 Halloween events from among the IDEDs above, because 
consecutive disturbances in them are caused by successive CMEs (\textit{e.g.} \opencite{Andrews2001}; 
\opencite{Gopalswamy2005}). As a result, 159 out of 656 IDEDs are excluded, and the remaining 
497 IDEDs are listed as candidates for ICME events. 

Next, we examine the relationship between CME/ICME pairs and selected IDEDs. In 
this examination, we use the list of near-Earth ICMEs and associated CMEs compiled by 
\inlinecite{Richardson2010}. This includes 322 ICMEs associated with a halo or a partial 
halo or normal CMEs during Solar Cycle 23; here, ``normal'' means that the exterior of 
CME is neither a halo nor a partial halo. In the above study, CMEs were observed by the 
SOHO/LASCO coronagraphs, and ICMEs were detected by \textit{in-situ} observation using spacecraft 
such as ACE and the \textit{Interplanetary Monitoring Platform}-8 (IMP-8). We compare the 
list of IDEDs with that of ICMEs using the assumption that an ICME caused the IDED. 
When an IDED is between the appearance date of an associated CME and the detection date 
of a near-Earth ICME, we assume that the IDED was related to the ICME. 

Using the above method, we find 66 IDEDs from our list that were probably related to 
ICMEs. However, we also find that 16 IDEDs of the 66 had multiple associated CMEs. For 
these 16 events, we identify the optimal one-to-one correspondence by comparing positions 
for LOS exhibiting high \textit{g}-values in a \textit{g}-map with the direction of the associated CME 
eruption in the LASCO field-of-view (FOV). 

At the end of this selection, we identify 50 CMEs and their associated ICMEs that were 
detected by the SOHO/LASCO, IPS, and \textit{in-situ} observations. For these, we estimate radial 
speeds and accelerations in interplanetary space using the method described in the next 
subsection.

\subsection{Estimations of ICME Radial Speeds and Accelerations} 
      \label{estimation}

The ICME radial speeds and accelerations are estimated in two interplanetary regions, 
\textit{i.e.} the region between SOHO and IPS observations (the SOHO--IPS region, from 0.1 to 
${\approx}$ 0.6 AU) and that between IPS and \textit{in-situ} observations (the IPS--Earth region, from ${\approx}$ 0.6 
to 1 AU). In these estimations, we assume that locations of LOS for disturbed sources in a 
\textit{g}-map give the location of the ICME. 

First, we calculate radial speeds at reference distances for each ICME. For each radio 
source of $g \ge 1.5$ in a \textit{g}-map, distances [${r_\mathrm{1}}$ and ${r_\mathrm{2}}$] and radial speeds 
[${v_\mathrm{1}}$ and ${v_\mathrm{2}}$] are derived from the following equations: 
\begin{equation}
  \label{eq.r1v1}
r_\mathrm{1} = \frac{r_\mathrm{S} + r_\mathrm{IPS}}{2},~
v_\mathrm{1} = \frac{r_\mathrm{IPS} - r_\mathrm{S}}{t_\mathrm{IPS} - T_\mathrm{SOHO}}~
\mbox{(for the SOHO--IPS region),}
\end{equation}
and
\begin{equation}
  \label{eq.r2v2}
r_\mathrm{2} = \frac{r_\mathrm{IPS} + r_\mathrm{E}}{2},~
v_\mathrm{2} = \frac{r_\mathrm{E} - r_\mathrm{IPS}}{T_\mathrm{Earth} - t_\mathrm{IPS}}~
\mbox{(for the IPS--Earth region),}
\end{equation}
respectively. Here, ${r_\mathrm{S}}$ is the minimum radius of SOHO/LASCO-C2 FOV, \textit{i.e.} 0.009 AU, 
${r_\mathrm{IPS}}$ is the radial distance of P-point on the LOS, ${r_\mathrm{E}}$ is the distance between 
the Sun and the Earth, \textit{i.e.} 1 AU, ${T_\mathrm{SOHO}}$ is the appearance time of CME in 
the SOHO/LASCO-C2 FOV, ${t_\mathrm{IPS}}$ is the observation time for a $g \ge 1.5$ source, and 
${T_\mathrm{Earth}}$ is the onset time of near-Earth ICME by \textit{in-situ} observation. 
Using these values, the average reference distances [${R_\mathrm{1}}$ and ${R_\mathrm{2}}$] and 
the average radial speeds [${V_\mathrm{1}}$ and ${V_\mathrm{2}}$] for the ICME are found 
for values of ${r_\mathrm{1}}$, ${r_\mathrm{2}}$, ${v_\mathrm{1}}$, and ${v_\mathrm{2}}$ 
for all $g \ge 1.5$ sources, respectively on a given day.

Next, we calculate accelerations using the values above. In these calculations, we use the 
approximation that the accelerations are constant within each region. The average accelerations, 
\textit{i.e.} ${a_\mathrm{1}}$ and ${a_\mathrm{2}}$, for ICMEs were given by 
\begin{equation}
  \label{eq.accel1} 
a_\mathrm{1} = \frac{1}{n} \sum_{k = 1}^{n} \frac{v_{\mathrm{IPS},k} - V_\mathrm{SOHO}}{t_{\mathrm{IPS},k} - T_\mathrm{SOHO}}~
\mbox{(for the SOHO--IPS region),}
\end{equation}
and 
\begin{equation}
  \label{eq.accel2} 
a_\mathrm{2} = \frac{1}{n} \sum_{k = 1}^{n} \frac{V_\mathrm{Earth} - v_{\mathrm{IPS},k}}{T_\mathrm{Earth} - t_{\mathrm{IPS},k}}~
\mbox{(for the IPS--Earth region),}
\end{equation}
respectively. Here, 
\begin{equation}
  \label{eq.vips}
v_{\mathrm{IPS},k} = \frac{v_{\mathrm{1},k} + v_{\mathrm{2},k}}{2}, 
\end{equation}
${t_{\mathrm{IPS},k}}$ is the observation time for each $g \ge 1.5$ source, $n$ is the number of $g \ge 1.5$ sources, and 
${V_\mathrm{SOHO}}$ and ${V_\mathrm{Earth}}$ are the radial speed of the CME and of the near-Earth ICME, respectively. 
For the value of ${V_\mathrm{SOHO}}$ in the halo or the partial halo CMEs, we use
\begin{equation}
  \label{eq.vsoho}
V_\mathrm{SOHO} = 1.20 \times V_\mathrm{POS}, 
\end{equation}
where ${V_\mathrm{POS}}$ is the speed measured in the sky plane by the SOHO/LASCO, because the 
coronagraph measurement for them tends to underestimate the radial speed \cite{Michalek2003}, 
while we use $V_\mathrm{SOHO} = V_\mathrm{POS}$ for the normal ones. 
In this study, we use the linear speeds reported in the SOHO/LASCO CME Catalog 
(\url{cdaw.gsfc.nasa.gov/CME_list/index.html}) for those of ${V_\mathrm{POS}}$ with a 0.08 AU reference distance 
corresponding to half the LASCO FOV value. Those are derived from the bright leading 
edges of CME \cite{Yashiro2004}, while the associated shocks show a faint structure ahead of 
them \cite{Ontiveros2009}, and then indicate the speeds of CME itself 
in the sky plane \cite{Vourlidas2012}. For values of ${V_\mathrm{Earth}}$, we use the average ICME 
speeds listed by \inlinecite{Richardson2010}. We note that the values of ${V_\mathrm{SOHO}}$ and ${V_\mathrm{Earth}}$ 
represent an average in the near-Sun and near-Earth regions, respectively, and ${V_\mathrm{1}}$, ${V_\mathrm{2}}$, 
${a_\mathrm{1}}$, and ${a_\mathrm{2}}$ are averages in the interplanetary space. The ICME speeds in the near-Earth region are 
measured when the spacecraft passes through them. Thus, those are equivalent to the plasma 
flow speed on the trajectory of the spacecraft during the passage of an ICME, indicated by 
the enhancement of the charge state and the rotation of magnetic-field direction \cite{Richardson2010}. 
The speed of the solar wind measured by \textit{in-situ} observations is sometimes 
highly variable during the passage of an ICME. However, the majority of ICMEs listed by 
them have only $< 100$ ${\mathrm{km~s^{-1}}}$ difference between the peak and average speeds. Hence, we 
consider it justified that the average flow speed can be used as the propagation speed of 
ICMEs.

\subsection{Classification of ICMEs} 
      \label{classification}

Here, we introduce ${V_\mathrm{IPS}}$ which is given as the average value of ${v_\mathrm{IPS}}$ for each ICME; 
the ${v_\mathrm{IPS}}$ is derived from Equation~\ref{eq.vips}. In addition, we also introduce ${V_\mathrm{bg}}$ 
as the speed of the background solar wind. To determine the value of ${V_\mathrm{bg}}$ as the average background 
wind speed between ${T_\mathrm{SOHO}}$ and ${T_\mathrm{Earth}}$ for each ICME, we used plasma data obtained by 
space-borne instruments including \textit{Solar Wind Electron, Proton, and Alpha Monitor} onboard ACE (ACE/SWEPAM: 
\opencite{McComas1998b}), \textit{Solar Wind Experiment} on \textit{Wind} (Wind/SWE: \opencite{Ogilvie1995}),
\textit{Massachusetts Institute of Technology Faraday cup experiment} on IMP-8 (IMP-8/MIT: \opencite{Bellomo1978}), 
and the \textit{Comprehensive Plasma Instrumentation} on GEOTAIL (GEOTAIL/CPI: \opencite{Frank1994}); 
these are determined from the NASA/GSFC OMNI dataset through OMNIWeb Plus (\url{omniweb.gsfc.nasa.gov/}). 

Using the values of ${V_\mathrm{SOHO}}$, ${V_\mathrm{IPS}}$, and ${V_\mathrm{bg}}$, we classify 
the 50 ICMEs into three types: fast ($V_{\mathrm{SOHO}} - V_\mathrm{bg} > 500$ ${\mathrm{km~s^{-1}}}$), 
moderate ($0$ $\mathrm{km~s^{-1}}$ $\le V_{\mathrm{SOHO}} - V_\mathrm{bg} \le 500$ ${\mathrm{km~s^{-1}}}$), and 
slow ($V_{\mathrm{SOHO}} - V_\mathrm{bg} < 0$ ${\mathrm{km~s^{-1}}}$). In our results, 
the numbers of fast, moderate, and slow ICMEs are 19, 25, and 6, respectively. Here, we eliminate 
5 of the 19 fast ICMEs and a moderate ICME because they show an extreme zigzag profile of propagation speeds, 
\textit{i.e.} $V_{\mathrm{1}} - V_{\mathrm{2}} > 1000$ ${\mathrm{km~s^{-1}}}$. 
The value of $V_{\mathrm{1}} - V_{\mathrm{2}} > 1000$ ${\mathrm{km~s^{-1}}}$ implies that the ICME has a 
strange acceleration, and then shows an unrealistic propagation. We also eliminate 4 of the 
24 moderate ICMEs and one of the six slow ones because they exhibit the unusual values 
of ${V_\mathrm{IPS}}$ of $V_\mathrm{IPS} - V_\mathrm{bg} > 500$ ${\mathrm{km~s^{-1}}}$ and 
$V_\mathrm{IPS} - V_\mathrm{bg} > 100$ ${\mathrm{km~s^{-1}}}$, respectively. The values 
of $V_\mathrm{IPS} - V_\mathrm{bg} > 500$ ${\mathrm{km~s^{-1}}}$ for moderate and 
$V_\mathrm{IPS} - V_\mathrm{bg} > 100$ ${\mathrm{km~s^{-1}}}$ for slow ICMEs 
imply that the ICME has a strange acceleration since ${V_\mathrm{IPS}}$ is larger than ${V_\mathrm{SOHO}}$ and 
${V_\mathrm{Earth}}$, and an unrealistic ICME propagation that indicates a higher speed in the region beyond 
coronagraph distances, and less at 1 AU. 

Finally, we obtain physical properties for 39 ICMEs which consist of 14 fast, 20 moderate, 
and five slow ones. 

\section{Results} 
      \label{results}

\subsection{Properties and Speed profiles of the 39 ICMEs}
      \label{icmeproperties}
      
\begin{figure}
      \begin{center}
      \centerline{\includegraphics[width=0.9\textwidth,clip=]{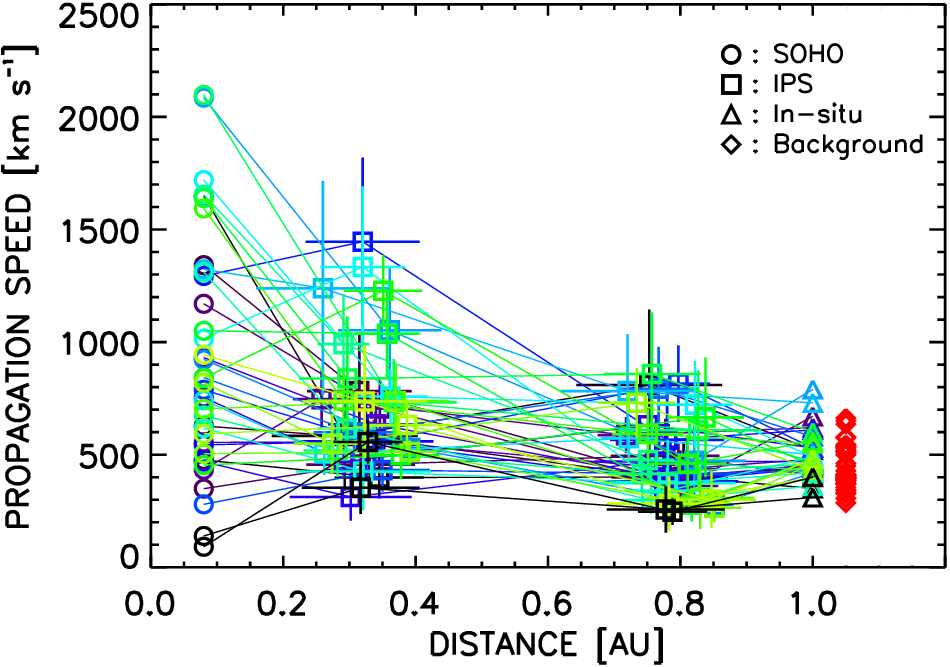}}
      \vspace{-0.05 \textwidth}
      \caption{
      Radial evolution of propagation speeds for the 39 ICMEs in this study. Open circle, 
      square, and triangle symbols indicate speeds of ICMEs measured by SOHO/LASCO, 
      IPS, and \textit{in-situ} observations, respectively. Symbols for each ICME are connected by solid 
      lines with the same color. Open diamond (red) symbols indicate speeds of the background solar 
      wind measured from \textit{in-situ} observations at 1 AU. 
      }
      \label{fig2}
      \end{center}
      \end{figure}

The properties of the 39 ICMEs identified from our analysis are listed in Tables \ref{table1} and \ref{table2} 
which including ${T_{\mathrm{IPS}}}$, ${R_\mathrm{0}}$, ${\alpha}$, ${\beta}$, and ${V_\mathrm{Tr}}$ 
in addition to ${T_{\mathrm{SOHO}}}$, ${V_{\mathrm{POS}}}$, ${V_{\mathrm{SOHO}}}$, ${R_\mathrm{1}}$, 
${V_\mathrm{1}}$, ${a_\mathrm{1}}$, ${R_\mathrm{2}}$, ${V_\mathrm{2}}$, ${a_\mathrm{2}}$, 
${T_{\mathrm{Earth}}}$, ${V_{\mathrm{Earth}}}$, and ${V_\mathrm{bg}}$ above. 
Here, ${T_{\mathrm{IPS}}}$ and ${R_\mathrm{0}}$ are the mean time and the average radial distance for 
an ICME detected by IPS observations; those are given as the averages of ${t_\mathrm{IPS}}$ and of 
${r_\mathrm{IPS}}$ for the $g \ge 1.5$ sources, respectively. The ${\alpha}$ and ${\beta}$ are 
the index and coefficient for a power-law form of the radial speed evolution described as 
\begin{equation}
  \label{eq.powerlaw}
V = {\beta}R^{\alpha}, 
\end{equation}
where $R$ is the heliocentric distance. ${V_\mathrm{Tr}}$ is the transit speed: 
\begin{equation}
  \label{eq.vtr}
V_\mathrm{Tr} = \frac{r_\mathrm{E}}{T_\mathrm{Earth} - T_\mathrm{SOHO}}. 
\end{equation}
This is equivalent to the average speed of ICMEs between the Sun and the Earth. 
In addition, we plot all of the speed profiles in order to show radial speed evolutions of 
ICMEs in Figure~\ref{fig2}. Here, data points for each ICME are connected by solid lines instead 
of fitting in Equation~(\ref{eq.powerlaw}). As shown here, ICME propagation speeds in 
the near-Sun region exhibit a wide range from 90 ${\mathrm{km~s^{-1}}}$ to 
${\approx}~2100$ ${\mathrm{km~s^{-1}}}$, while those in the near-Earth 
region range from 310 ${\mathrm{km~s^{-1}}}$ to 790 ${\mathrm{km~s^{-1}}}$. 
Moreover, the range of ICME propagation speeds in interplanetary space decreases 
with increasing distance. In addition, speeds of the background solar wind also show 
a relatively narrow span from 286 ${\mathrm{km~s^{-1}}}$ to 662 ${\mathrm{km~s^{-1}}}$. 

     %
     \captionwidth=17.2cm
     \begin{landscape}
      \begin{table}
      \caption{Properties derived from SOHO/LASCO observations and those in the SOHO--IPS region derived from IPS observations for 39 ICMEs }
      \label{table1}
      \scalebox{0.90}{
      \begin{tabular}[!p]{cccccccccccccccccc}
\multicolumn{6}{l}{\small{during 1997\,--\,2009.}}  & ~~~ \\
      \hline
No. & \multicolumn{6}{c}{SOHO/LASCO} &  & \multicolumn{10}{c}{IPS} \\ \cline{2-7} \cline{9-18}
~~~ & ~~~~~~~~~~~ & ~~~~ & ~~~ & ~~~ & ~~~ & ~~~ &  & ~~~~~~~~~~~ & \multicolumn{3}{c}{Disturbance} & \multicolumn{6}{c}{SOHO--IPS region} \\
~~~ & Date & Time & ${V_\mathrm{POS}}$ & ${V_\mathrm{SOHO}}$ & CME & PA &  & Date & Time & \multicolumn{2}{c}{${R_\mathrm{0}}$~[AU]} & \multicolumn{2}{c}{$R_\mathrm{1}$~[AU]} & \multicolumn{2}{c}{$V_\mathrm{1}$~[$\mathrm{km~s^{-1}}$]} & \multicolumn{2}{c}{$a_\mathrm{1}$~[$\mathrm{m~s^{-2}}$]} \\
~~~ & [ddmmmyy] & [hhmm] & [${\mathrm{km~s^{-1}}}$] & [${\mathrm{km~s^{-1}}}$] & Type & [deg] &  & [ddmmmyy] & [hhmm] & Aver. & ${\sigma}$ & Aver. & ${\sigma}$ & Aver. & ${\sigma}$ & Aver. & ${\sigma}$ \\
      \hline
~1 & 06~Dec~1997 & 1027 & ~397 & ~476 & PH & $317$ &  & 09~Dec~1997 & 0115 & 0.61 & 0.13 & 0.35 & 0.07 & ~399 & ~93 & $~-0.34$ & 0.36 \\
~2 & 29~Apr~1998 & 1658 & 1374 & 1649 & FH & $-99$ &  & 01~May~1998 & 0358 & 0.51 & 0.22 & 0.29 & 0.11 & ~584 & 243 & $~-7.55$ & 1.47 \\
~3 & 04~Nov~1998 & 0754 & ~523 & ~628 & FH & $-99$ &  & 06~Nov~1998 & 0259 & 0.60 & 0.10 & 0.34 & 0.05 & ~570 & 116 & $~-0.95$ & 0.45 \\
~4 & 05~Nov~1998 & 2044 & 1118 & 1342 & FH & $-99$ &  & 07~Nov~1998 & 0143 & 0.55 & 0.16 & 0.32 & 0.08 & ~782 & 257 & $~-7.20$ & 1.36 \\
~5 & 13~Apr~1999 & 0330 & ~291 & ~349 & PH & $228$ &  & 15~Apr~1999 & 0453 & 0.55 & 0.16 & 0.32 & 0.08 & ~456 & 130 & $~~0.71$ & 0.55 \\
~6 & 24~Jun~1999 & 1331 & ~975 & 1170 & FH & $-99$ &  & 26~Jun~1999 & 0114 & 0.61 & 0.10 & 0.35 & 0.05 & ~705 & 135 & $~-4.96$ & 0.61 \\
~7 & 28~Jul~1999 & 0430 & ~361 & ~433 & PH & $112$ &  & 29~Jul~1999 & 0408 & 0.44 & 0.09 & 0.26 & 0.04 & ~747 & 135 & $~~2.74$ & 0.79 \\
~8 & 28~Jul~1999 & 0906 & ~462 & ~554 & FH & $-99$ &  & 30~Jul~1999 & 0438 & 0.60 & 0.12 & 0.34 & 0.06 & ~557 & ~97 & $~-0.37$ & 0.43 \\
~9 & 17~Aug~1999 & 1331 & ~776 & ~931 & PH & $~61$ &  & 19~Aug~1999 & 0403 & 0.60 & 0.14 & 0.34 & 0.07 & ~635 & 138 & $~-3.04$ & 0.67 \\
10 & 21~May~2000 & 0726 & ~629 & ~755 & PH & $304$ &  & 23~May~2000 & 0311 & 0.50 & 0.07 & 0.29 & 0.03 & ~469 & ~60 & $~-1.31$ & 0.29 \\
11 & 31~May~2000 & 0806 & ~391 & ~469 & PH & $~46$ &  & 03~Jun~2000 & 0417 & 0.52 & 0.18 & 0.30 & 0.09 & ~312 & 105 & $~-0.32$ & 0.37 \\
12 & 07~Jul~2000 & 1026 & ~453 & ~544 & FH & $-99$ &  & 09~Jul~2000 & 0345 & 0.57 & 0.20 & 0.33 & 0.10 & ~559 & 180 & $~-0.49$ & 0.80 \\
13 & 11~Jul~2000 & 1327 & 1078 & 1294 & FH & $-99$ &  & 12~Jul~2000 & 0459 & 0.56 & 0.17 & 0.32 & 0.09 & 1446 & 373 & $~-5.27$ & 3.60 \\
14 & 17~Jul~2000 & 0854 & ~788 & ~788 & NM & $~95$ &  & 19~Jul~2000 & 0421 & 0.59 & 0.10 & 0.34 & 0.05 & ~557 & ~78 & $~-0.65$ & 0.56 \\
15 & 06~Aug~2000 & 1830 & ~233 & ~280 & PH & $105$ &  & 09~Aug~2000 & 0503 & 0.62 & 0.14 & 0.35 & 0.07 & ~432 & ~95 & $~~0.68$ & 0.40 \\
16 & 09~Aug~2000 & 1630 & ~702 & ~842 & FH & $-99$ &  & 11~Aug~2000 & 0446 & 0.54 & 0.12 & 0.31 & 0.06 & ~600 & 123 & $~-1.12$ & 0.78 \\
17 & 29~Aug~2000 & 1830 & ~769 & ~923 & PH & $347$ &  & 01~Sep~2000 & 0106 & 0.70 & 0.06 & 0.39 & 0.03 & ~530 & ~57 & $~-2.65$ & 0.19 \\
18 & 08~Nov~2000 & 2306 & 1738 & 2086 & PH & $271$ &  & 10~Nov~2000 & 0017 & 0.64 & 0.16 & 0.36 & 0.08 & 1053 & 283 & $-14.60$ & 1.88 \\
19 & 11~Apr~2001 & 1331 & 1103 & 1324 & FH & $-99$ &  & 12~Apr~2001 & 0334 & 0.44 & 0.20 & 0.26 & 0.10 & 1239 & 477 & $~-6.38$ & 4.77 \\
20 & 14~Aug~2001 & 1601 & ~618 & ~742 & FH & $-99$ &  & 16~Aug~2001 & 0341 & 0.44 & 0.08 & 0.26 & 0.04 & ~507 & ~97 & $~-1.57$ & 0.45 \\
     \hline
      \end{tabular}
      }
      \end{table}
    \end{landscape}
     %
     \captionwidth=17.2cm
     \begin{landscape}
      \begin{table}
      \scalebox{0.90}{
      \begin{tabular}[!p]{cccccccccccccccccc}
\multicolumn{6}{l}{\small{(Continued from the previous page)}}  & ~~~ \\
      \hline
No. & \multicolumn{6}{c}{SOHO/LASCO} &  & \multicolumn{10}{c}{IPS} \\ \cline{2-7} \cline{9-18}
~~~ & ~~~~ & ~~~~ & ~~~ & ~~~ & ~~~ & ~~ &  & ~~~~ & \multicolumn{2}{c}{Disturbance} & \multicolumn{6}{c}{SOHO--IPS region} \\
~~~ & Date & Time & ${V_\mathrm{POS}}$ & ${V_\mathrm{SOHO}}$ & CME & PA &  & Date & Time & \multicolumn{2}{c}{${R_\mathrm{0}}$~[AU]} & \multicolumn{2}{c}{${R_\mathrm{1}}$~[AU]} & \multicolumn{2}{c}{${V_\mathrm{1}}$~[${\mathrm{km~s^{-1}}}$]} & \multicolumn{2}{c}{${a_\mathrm{1}}$~[${\mathrm{m~s^{-2}}}$]} \\
~~~ & [ddmmmyy] & [hhmm] & [${\mathrm{km~s^{-1}}}$] & [${\mathrm{km~s^{-1}}}$] & Type & [deg] &  & [ddmmmyy] & [hhmm] & Aver. & ${\sigma}$ & Aver. & ${\sigma}$ & Aver. & ${\sigma}$ & Aver. & ${\sigma}$ \\
     \hline
21 & 25~Aug~2001 & 1650 & 1433 & 1720 & FH & $-99$ &  & 27~Aug~2001 & 0337 & 0.64 & 0.11 & 0.36 & 0.06 & ~758 & 122 & $~-7.85$ & 0.85 \\
22 & 28~Sep~2001 & 0854 & ~846 & 1015 & FH & $-99$ &  & 29~Sep~2001 & 0220 & 0.56 & 0.13 & 0.32 & 0.06 & 1334 & 359 & $~-2.72$ & 2.84 \\
23 & 22~Oct~2001 & 1826 & ~618 & ~618 & NM & $131$ &  & 25~Oct~2001 & 0116 & 0.56 & 0.20 & 0.32 & 0.10 & ~425 & 171 & $~-1.11$ & 0.58 \\
24 & 25~Oct~2001 & 1526 & 1092 & 1310 & FH & $-99$ &  & 27~Oct~2001 & 0137 & 0.51 & 0.09 & 0.30 & 0.04 & ~617 & 136 & $~-6.94$ & 0.55 \\
25 & 17~Nov~2001 & 0530 & 1379 & 1655 & FH & $-99$ &  & 18~Nov~2001 & 0229 & 0.50 & 0.08 & 0.29 & 0.04 & ~991 & 222 & $-12.23$ & 1.45 \\
26 & 29~Jul~2002 & 1207 & ~562 & ~674 & PH & $~13$ &  & 31~Jul~2002 & 0212 & 0.48 & 0.10 & 0.28 & 0.05 & ~519 & 117 & $~-1.52$ & 0.47 \\
27 & 05~Sep~2002 & 1654 & 1748 & 2098 & FH & $-99$ &  & 07~Sep~2002 & 0503 & 0.65 & 0.19 & 0.37 & 0.10 & ~733 & 192 & $-10.98$ & 1.13 \\
28 & 28~May~2003 & 0050 & 1366 & 1639 & FH & $-99$ &  & 29~May~2003 & 0210 & 0.51 & 0.15 & 0.30 & 0.07 & ~838 & 275 & $~-8.72$ & 2.05 \\
29 & 14~Jun~2003 & 0154 & ~875 & 1050 & PH & $~26$ &  & 15~Jun~2003 & 0311 & 0.63 & 0.10 & 0.36 & 0.05 & 1038 & 194 & $~-4.30$ & 1.10 \\
30 & 14~Aug~2003 & 2006 & ~378 & ~454 & FH & $-99$ &  & 17~Aug~2003 & 0409 & 0.68 & 0.12 & 0.38 & 0.06 & ~497 & 106 & $~~0.62$ & 0.63 \\
31 & 22~Jul~2004 & 0731 & ~700 & ~840 & PH & $~66$ &  & 23~Jul~2004 & 0406 & 0.62 & 0.12 & 0.35 & 0.06 & 1228 & 158 & $~~0.06$ & 1.28 \\
32 & 12~Sep~2004 & 0036 & 1328 & 1594 & FH & $-99$ &  & 13~Sep~2004 & 0405 & 0.50 & 0.11 & 0.29 & 0.05 & ~741 & 138 & $~-9.40$ & 0.84 \\
33 & 26~May~2005 & 1506 & ~586 & ~703 & FH & $-99$ &  & 28~May~2005 & 0334 & 0.66 & 0.16 & 0.37 & 0.08 & ~738 & 161 & $~-1.38$ & 0.75 \\
34 & 26~May~2005 & 2126 & ~420 & ~504 & PH & $144$ &  & 29~May~2005 & 0522 & 0.70 & 0.08 & 0.39 & 0.04 & ~513 & ~49 & $~-0.57$ & 0.18 \\
35 & 07~Jul~2005 & 1706 & ~683 & ~820 & FH & $-99$ &  & 09~Jul~2005 & 0347 & 0.47 & 0.12 & 0.27 & 0.06 & ~549 & 132 & $~-1.44$ & 0.74 \\
36 & 05~Aug~2005 & 0854 & ~494 & ~593 & PH & $~23$ &  & 07~Aug~2005 & 0547 & 0.69 & 0.13 & 0.39 & 0.07 & ~632 & 115 & $~-0.78$ & 0.48 \\
37 & 26~Aug~2006 & 2057 & ~786 & ~943 & PH & $164$ &  & 28~Aug~2006 & 0425 & 0.57 & 0.19 & 0.32 & 0.09 & ~736 & 250 & $~-3.83$ & 1.17 \\
38 & 12~Sep~2008 & 1030 & ~~91 & ~~91 & NM & $~89$ &  & 14~Sep~2008 & 0449 & 0.58 & 0.10 & 0.33 & 0.05 & ~556 & ~96 & $~~2.04$ & 0.34 \\
39 & 29~May~2009 & 0930 & ~139 & ~139 & NM & $258$ &  & 01~Jun~2009 & 0148 & 0.56 & 0.18 & 0.32 & 0.09 & ~353 & 116 & $~~0.72$ & 0.32 \\
     \hline
\multicolumn{18}{p{21.4cm}}{
Column: (1) Event number; (2)\,--\,(3) Appearance date [ddmmmyy] and time [hhmm] of 
an ICME--associated CME observed by SOHO/LASCO;(4) Speed in the sky plane measured by SOHO/LASCO with 0.08 AU 
of reference distance; (5) Radial speed estimated using $V_{\mathrm{SOHO}} = 1.20 \times V_{\mathrm{POS}}$; (6) Type of CME [FH, PH, and 
NM mean Full Halo, Partial Halo, and Normal CME, respectively]; 
(7) Position angle measured from solar North in degrees (counter-clockwise)[$- 99$ means Full Halo]; (8)\,--\,(9) Observation date [ddmmmyy] and 
mean time [hhmm] of IPS disturbance event day ; (10)\,--\,(11) Average and standard errors for the distance of observed disturbance [${R_\mathrm{0}}$]; 
(12)\,--\,(13) Average and standard error for the reference distance [${R_{1}}$] (in the SOHO--IPS region); (14)\,--\,(15) Average and standard 
error for the speed [${V_\mathrm{1}}$] (in the SOHO--IPS region); (16)\,--\,(17) Average and standard error for acceleration [${a_\mathrm{1}}$] 
(in the SOHO--IPS region).
      }
      \end{tabular}
      }
      \end{table}
     \end{landscape}
     %
     \captionwidth=17.2cm
     \begin{landscape}
      \begin{table}
      \caption{Properties in the IPS--Earth region derived from IPS observations, detection dates, times, and speeds obtained by \textit{in-situ} }
      \label{table2}
      \scalebox{0.90}{
      \begin{tabular}{ccccccccccccccccc}
\multicolumn{14}{l}{\small{observations at 1 AU, fitting parameters and speeds of the background solar wind for 39 ICMEs during 1997\,--\,2009.}}  & ~~~ \\
      \hline
No. & \multicolumn{6}{c}{IPS} &  & \multicolumn{3}{c}{\textit{in~situ}} & \multicolumn{2}{c}{Parameters for} & ~~~ & ~~~ &~\\ \cline{2-7} \cline{9-11}
~~~ & \multicolumn{6}{c}{IPS--Earth region} & & ~~~ & ~~~ & ~~~ & \multicolumn{2}{c}{power-law equation} & \multicolumn{3}{p{3.7cm}}{~~~~~~~~~~~~~~~Background wind} \\
~~~ & \multicolumn{2}{c}{$R_\mathrm{2}$~[AU]} &\multicolumn{2}{c}{$V_\mathrm{2}$~[$\mathrm{km~s^{-1}}$]} & \multicolumn{2}{c}{$a_\mathrm{2}$~[$\mathrm{m~s^{-2}}$]} & & Date & Time & $V_\mathrm{Earth}$ & Index & Coefficient & $V_\mathrm{Tr}$ & \multicolumn{2}{p{2.2cm}}{~$V_\mathrm{bg}$~[$\mathrm{km~s^{-1}}$]} \\
~~~ & Aver. & ${\sigma}$ & Aver. & ${\sigma}$ & Aver. & ${\sigma}$ &  & [ddmmmyyyy] & [hhmm] & [$\mathrm{km~s^{-1}}$] & ${\alpha}$ & ${\beta}$ & [$\mathrm{km~s^{-1}}$] & Aver. & $\sigma$ \\
      \hline
~1 & 0.81 & 0.07 & 401 & 153 & $-0.35$ & 0.56 & & 10~Dec~1997 & 1800 & 350 & $-0.102$ & 366.9 & 401 & 354 &~24 \\
~2 & 0.75 & 0.11 & 809 & 335 & $-1.95$ & 2.06 & & 02~May~1998 & 0500 & 520 & $-0.374$ & 547.2 & 692 & 369 &~47 \\
~3 & 0.80 & 0.05 & 391 & 105 & $-0.20$ & 0.45 & & 07~Nov~1998 & 2200 & 450 & $-0.167$ & 426.9 & 482 & 385 &~27 \\
~4 & 0.78 & 0.08 & 399 & 145 & $-0.82$ & 0.83 & & 09~Nov~1998 & 0100 & 450 & $-0.478$ & 411.5 & 544 & 385 &~27 \\
~5 & 0.78 & 0.08 & 495 & 165 & $-0.49$ & 0.73 & & 16~Apr~1999 & 1800 & 410 & $~0.094$ & 465.0 & 480 & 398 &~15 \\
~6 & 0.81 & 0.05 & 361 & 101 & $~0.85$ & 0.48 & & 27~Jun~1999 & 2200 & 670 & $-0.340$ & 483.9 & 516 & 306 &~31 \\
~7 & 0.72 & 0.04 & 587 & ~80 & $-0.33$ & 0.47 & & 30~Jul~1999 & 2000 & 620 & $~0.111$ & 658.4 & 654 & 377 &~43 \\
~8 & 0.80 & 0.06 & 436 & 112 & $-0.12$ & 0.49 & & 31~Jul~1999 & 1900 & 480 & $-0.080$ & 466.9 & 507 & 545 &~22 \\
~9 & 0.80 & 0.07 & 385 & 134 & $-0.32$ & 0.60 & & 20~Aug~1999 & 2300 & 460 & $-0.325$ & 416.8 & 510 & 635 &~67 \\
10 & 0.75 & 0.03 & 630 & ~78 & $-0.17$ & 0.38 & & 24~May~2000 & 1200 & 530 & $-0.098$ & 530.8 & 542 & 579 &~~7 \\
11 & 0.76 & 0.09 & 470 & 173 & $~0.53$ & 0.60 & & 04~Jun~2000 & 2200 & 470 & $~0.022$ & 433.4 & 378 & 462 &~55 \\
12 & 0.79 & 0.10 & 381 & 177 & $-0.18$ & 0.71 & & 11~Jul~2000 & 0200 & 440 & $-0.120$ & 423.0 & 474 & 371 &~13 \\
13 & 0.78 & 0.09 & 566 & 199 & $-3.46$ & 1.75 & & 13~Jul~2000 & 1300 & 610 & $-0.347$ & 638.2 & 874 & 500 &~24 \\
14 & 0.80 & 0.05 & 816 & 170 & $-2.12$ & 1.18 & & 20~Jul~2000 & 0100 & 530 & $-0.079$ & 611.8 & 648 & 574 &~43 \\
15 & 0.81 & 0.07 & 414 & 149 & $~0.06$ & 0.62 & & 10~Aug~2000 & 1900 & 430 & $~0.165$ & 447.8 & 430 & 412 &~36 \\
16 & 0.77 & 0.06 & 793 & 186 & $-1.33$ & 1.17 & & 12~Aug~2000 & 0500 & 580 & $-0.090$ & 634.9 & 686 & 412 &~36 \\
17 & 0.85 & 0.03 & 276 & ~59 & $~0.11$ & 0.23 & & 02~Sep~2000 & 2200 & 420 & $-0.398$ & 340.2 & 417 & 529 &~47 \\
18 & 0.82 & 0.08 & 473 & 218 & $~0.25$ & 1.49 & & 11~Nov~2000 & 0800 & 790 & $-0.492$ & 600.6 & 730 & 650 &184 \\
19 & 0.72 & 0.10 & 782 & 253 & $-2.69$ & 2.28 & & 13~Apr~2001 & 0900 & 730 & $-0.253$ & 753.9 & 955 & 537 &~26 \\
20 & 0.72 & 0.04 & 573 & ~85 & $-0.27$ & 0.40 & & 17~Aug~2001 & 2000 & 500 & $-0.125$ & 502.2 & 546 & 395 &~43 \\
     \hline
      \end{tabular}
      }
      \end{table}
    \end{landscape}
     %
     \captionwidth=17.2cm
     \begin{landscape}
      \begin{table}
      \scalebox{0.90}{
      \begin{tabular}{ccccccccccccccccc}
\multicolumn{6}{l}{\small{(Continued from the previous page)}}  & ~~~ \\
      \hline
No. & \multicolumn{6}{c}{IPS} &  & \multicolumn{3}{c}{\textit{in~situ}} & \multicolumn{2}{c}{Parameters for} & ~~~ & ~~~ &~\\ \cline{2-7} \cline{9-11}
~~~ & \multicolumn{6}{c}{IPS--Earth region} & & ~~~ & ~~~ & ~~~ & \multicolumn{2}{c}{power-law equation} & \multicolumn{3}{p{3.7cm}}{~~~~~~~~~~~~~~~Background wind} \\
~~~ & \multicolumn{2}{c}{$R_\mathrm{2}$~[AU]} &\multicolumn{2}{c}{$V_\mathrm{2}$~[$\mathrm{km~s^{-1}}$]} & \multicolumn{2}{c}{$a_\mathrm{2}$~[$\mathrm{m~s^{-2}}$]} & & Date & Time & $V_\mathrm{Earth}$ & Index & Coefficient & $V_\mathrm{Tr}$ & \multicolumn{2}{p{2.2cm}}{~$V_\mathrm{bg}$~[$\mathrm{km~s^{-1}}$]} \\
~~~ & Aver. & ${\sigma}$ & Aver. & ${\sigma}$ & Aver. & ${\sigma}$ &  & [ddmmmyyyy] & [hhmm] & [$\mathrm{km~s^{-1}}$] & ${\alpha}$ & ${\beta}$ & [$\mathrm{km~s^{-1}}$] & Aver. & $\sigma$ \\
      \hline
21 & 0.82 & 0.06 & 722 & 194 & $-3.43$ & 1.44 & & 28~Aug~2001 & 0000 & 490 & $-0.441$ & 545.6 & 752 & 410 &~21 \\
22 & 0.78 & 0.06 & 341 & 100 & $-1.78$ & 0.92 & & 01~Oct~2001 & 0800 & 490 & $-0.397$ & 467.5 & 584 & 513 &~33 \\
23 & 0.78 & 0.10 & 374 & 192 & $~0.11$ & 0.64 & & 27~Oct~2001 & 0300 & 420 & $-0.175$ & 379.7 & 397 & 393 &~32 \\
24 & 0.76 & 0.04 & 298 & ~61 & $-0.39$ & 0.28 & & 29~Oct~2001 & 2200 & 360 & $-0.569$ & 306.2 & 405 & 393 &~32 \\
25 & 0.75 & 0.04 & 477 & ~83 & $-1.93$ & 0.70 & & 19~Nov~2001 & 2200 & 430 & $-0.553$ & 435.4 & 644 & 399 &~20 \\
26 & 0.74 & 0.05 & 414 & ~81 & $-0.03$ & 0.34 & & 02~Aug~2002 & 0600 & 460 & $-0.175$ & 424.9 & 462 & 428 &~34 \\
27 & 0.83 & 0.10 & 609 & 266 & $-2.42$ & 1.78 & & 08~Sep~2002 & 0400 & 470 & $-0.562$ & 482.9 & 703 & 400 &~22 \\
28 & 0.76 & 0.07 & 858 & 274 & $-2.91$ & 2.19 & & 30~May~2003 & 0200 & 600 & $-0.344$ & 649.3 & 844 & 662 &~32 \\
29 & 0.82 & 0.05 & 277 & ~73 & $-0.89$ & 0.51 & & 17~Jun~2003 & 1000 & 480 & $-0.434$ & 410.4 & 518 & 520 &~44 \\
30 & 0.84 & 0.06 & 666 & 265 & $-1.84$ & 1.70 & & 18~Aug~2003 & 0100 & 450 & $~0.068$ & 543.0 & 540 & 534 &~55 \\
31 & 0.81 & 0.06 & 460 & 128 & $-2.33$ & 0.78 & & 24~Jul~2004 & 1400 & 560 & $-0.220$ & 583.3 & 762 & 450 &~61 \\
32 & 0.75 & 0.05 & 591 & 114 & $-0.93$ & 0.66 & & 14~Sep~2004 & 1500 & 550 & $-0.417$ & 516.0 & 666 & 286 &~37 \\
33 & 0.83 & 0.08 & 307 & 137 & $-0.38$ & 0.60 & & 30~May~2005 & 0100 & 460 & $-0.254$ & 411.0 & 507 & 342 &~56 \\
34 & 0.85 & 0.04 & 264 & ~65 & $~0.43$ & 0.22 & & 31~May~2005 & 0400 & 460 & $-0.141$ & 370.6 & 405 & 342 &~56 \\
35 & 0.73 & 0.06 & 731 & 154 & $-1.93$ & 0.85 & & 10~Jul~2005 & 1000 & 430 & $-0.166$ & 516.8 & 640 & 332 &~11 \\
36 & 0.85 & 0.07 & 300 & 127 & $~0.09$ & 0.51 & & 09~Aug~2005 & 0000 & 480 & $-0.175$ & 410.9 & 477 & 537 &~82 \\
37 & 0.78 & 0.09 & 283 & 120 & $-0.48$ & 0.58 & & 30~Aug~2006 & 2000 & 400 & $-0.429$ & 348.5 & 437 & 511 &~83 \\
38 & 0.79 & 0.05 & 247 & ~58 & $-0.01$ & 0.20 & & 17~Sep~2008 & 0400 & 400 & $~0.486$ & 425.8 & 366 & 406 &107 \\
39 & 0.78 & 0.09 & 256 & 103 & $~0.02$ & 0.28 & & 04~Jun~2009 & 0200 & 310 & $~0.276$ & 327.7 & 304 & 327 &~26 \\
     \hline
\multicolumn{16}{p{19.7cm}}{
Column: (1) Event number [identical with column (1) in Table \ref{table1}]; 
(2)\,--\,(3) Average and standard error for the reference distance [$R_\mathrm{2}$] (in the IPS--Earth region); 
(4)\,--\,(5) Average and standard error for the speed [$V_\mathrm{2}$] (in the IPS--Earth region); 
(6)\,--\,(7) Average and standard error for the acceleration [$a_\mathrm{2}$] (in the IPS--Earth region); 
(8)\,--\,(9) Detection date [ddmmmyy] and time [hhmm] of a near-Earth ICME by \textit{in-situ} observation at 1 AU; 
(10) Near-Earth ICME speed measured by \textit{in-situ} observation at 1 AU; 
(11)\,--\,(12); Index [${\alpha}$] and coefficient [${\beta}$] for a power-law form of radial speed evolution; 
(13) 1 AU transit speed [${V_\mathrm{Tr}}$] derived from the CME appearance and the ICME detection; 
(14)\,--\,(15) Average and standard deviation for the speed of the background wind [$V_\mathrm{bg}$] measured by 
spacecraft including ACE, \textit{Wind}, IMP-8, and \textit{GEOTAIL}.
} & ~\\
      \end{tabular}
      }
      \end{table}
    \end{landscape}
      %

\subsection{Fast, Moderate, and Slow ICMEs, and Their Accelerations}
      \label{accelerations}
           
\begin{figure}
      \begin{center}
      \centerline{\includegraphics[width=0.8\textwidth,clip=]{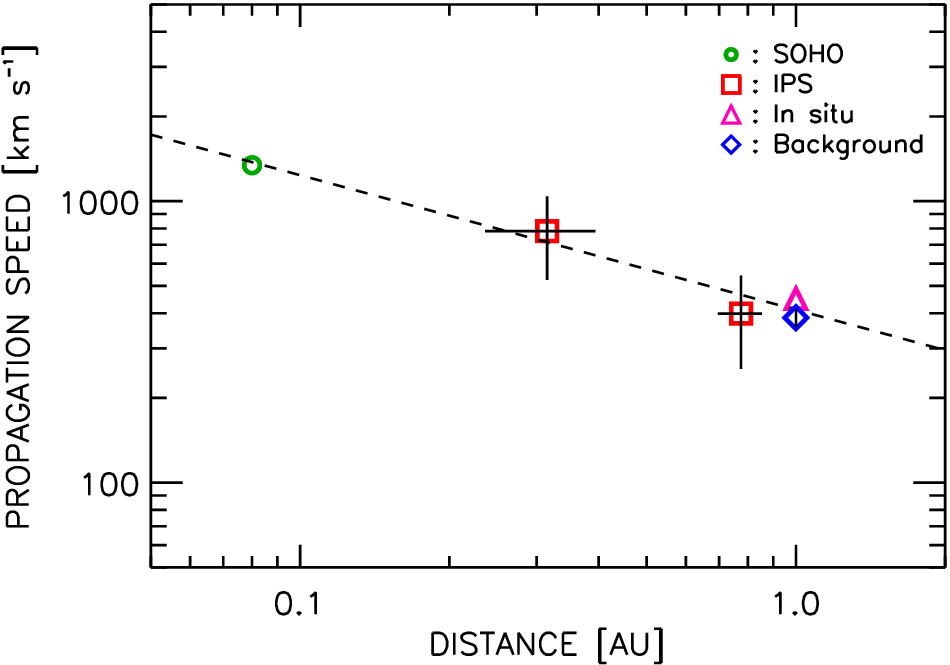}}
      \vspace{-0.05 \textwidth}
      \caption{
      Speed profile for the ICME event between 5 and 9 November 1998. This is an example of a fast 
      ICME. In this event, IPS disturbance event day is 7 November 1998. Open circle, square, and triangle denote 
      measurements of ICME speed from SOHO/LASCO, IPS, and \textit{in-situ} observations, respectively. An open 
      diamond indicates the speed of the background solar wind measured by \textit{in-situ} observations, and the dashed 
      line represents the power-law fit to the data using Equation~(\ref{eq.powerlaw}). Horizontal and vertical error bars are also 
      plotted using ${\sigma}$ values (standard error) for the reference distances [${R_\mathrm{1}}$ and ${R_\mathrm{2}}$] and 
      those for the speeds [${V_\mathrm{1}}$, ${V_\mathrm{2}}$, and ${V_\mathrm{bg}}$].
      }
      \label{fig3}
      \end{center}
      \end{figure}

\begin{figure}
      \begin{center}
      \centerline{\includegraphics[width=0.8\textwidth,clip=]{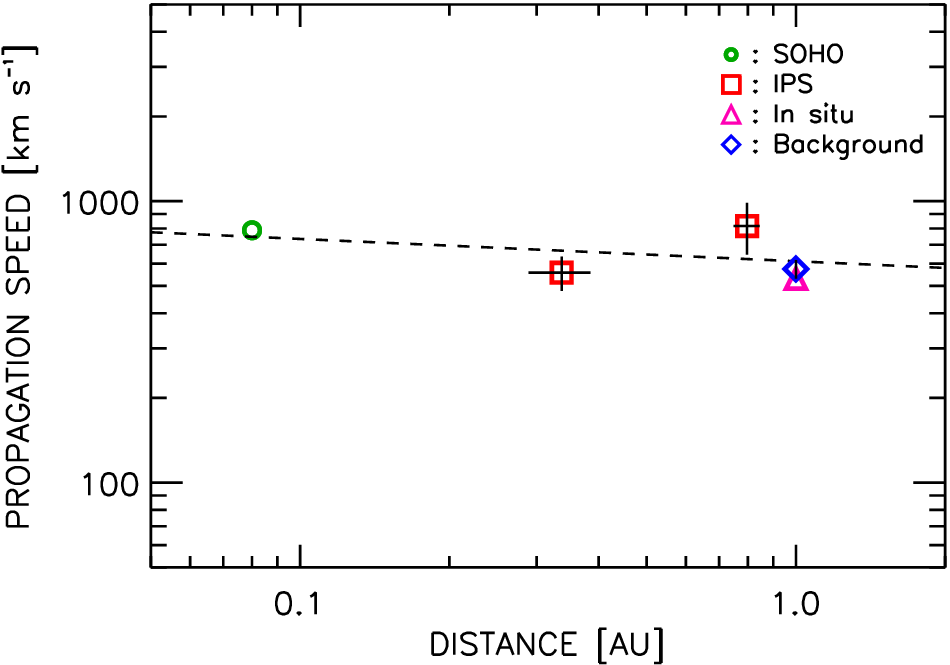}}
      \vspace{-0.05 \textwidth}
      \caption{
      Speed profile for the ICME event between 17 and 20 July 2000. This is an example of a moderate ICME. In this event, 
      IPS disturbance event day is 19 July 2000. Open circle, square, and triangle denote measurements of ICME speed 
      from SOHO/LASCO, IPS, and \textit{in-situ} observations, respectively. An open diamond indicates the speed of the background solar 
      wind measured by \textit{in-situ} observations, and a dashed line represents the power-law fit to the data using Equation~(\ref{eq.powerlaw}). 
      }
      \label{fig4}
      \end{center}
      \end{figure}

\begin{figure}
      \begin{center}
      \centerline{\includegraphics[width=0.8\textwidth,clip=]{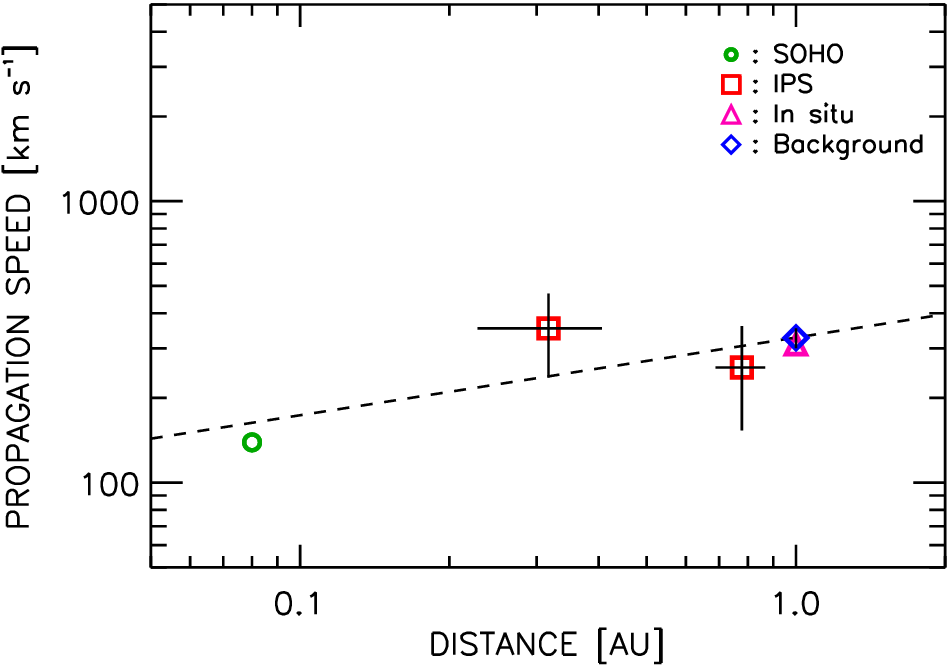}}
      \vspace{-0.05 \textwidth}
      \caption{
      Speed profile for the ICME event between 29 May and 4 June 2009. This is an example of a slow ICME. In this event, 
      IPS disturbance event day is 1 June 2009. Open circle, square, and triangle denote measurements of ICME speed 
      from SOHO/LASCO, IPS, and \textit{in-situ} observations, respectively. An open diamond indicates the speed of the background solar 
      wind measured by \textit{in-situ} observations, and a dashed line represents the power-law fit to the data using Equation~(\ref{eq.powerlaw}). 
      }
      \label{fig5}
      \end{center}
      \end{figure}

For the fast, moderate, and slow ICMEs, we show representative examples of speed profiles 
in Figures~\ref{fig3}, \ref{fig4}, and \ref{fig5}, respectively. These are plotted using 
the values of ${V_\mathrm{SOHO}}$, ${R_\mathrm{1}}$, ${V_\mathrm{1}}$, ${R_\mathrm{2}}$, 
${V_\mathrm{2}}$, ${V_\mathrm{Earth}}$, and ${V_\mathrm{bg}}$. Figure \ref{fig3} shows 
a speed profile for a fast ICME observed as a halo by SOHO/LASCO on 5 November 1998, 
a subsequent disturbance from the IPS observations on 7 November 1998, and the event 
detected at 1 AU by \textit{in-situ} observations on 9 November 1998 (see No. 4 in Tables \ref{table1} and \ref{table2}). 
These data show that the ICME speed rapidly decreases to the value of ${V_\mathrm{bg}}$ with 
an increase in radial distance; the initial speed ${V_\mathrm{SOHO}}$ value is 1342 
${\mathrm{km~s^{-1}}}$, while $V_\mathrm{bg} = 385$ ${\mathrm{km~s^{-1}}}$ for this ICME. 
This speed profile is well fit by a power-law function; the fitting-line has a value of 
${\alpha} = -0.478$ from Equation (\ref{eq.powerlaw}). Figure \ref{fig4} shows the speed 
profile for a moderate ICME; this ICME was observed as a normal event 
(neither a halo nor a partial halo) by SOHO/LASCO on 17 July 2000, on 19 July 2000 in IPS, 
and detected by \textit{in-situ} observations on 20 July 2000 (see No. 14 in Tables \ref{table1} and \ref{table2}). 
As shown here, for this ICME, the 788 ${\mathrm{km~s^{-1}}}$ initial speed gradually 
decreases to $V_\mathrm{bg} = 574$ ${\mathrm{km~s^{-1}}}$ with an increase in radial distance; 
we have a value of ${\alpha} = -0.079$. Figure \ref{fig5} exhibits a speed profile for a slow 
ICME observed as a normal event by SOHO/LASCO on 29 May 2009, on 1 June 2009 by IPS observations, 
and detected by \textit{in-situ} observations on 4 June 2009 (see No. 39 in Tables \ref{table1} and \ref{table2}). 
For this event, we confirm that $V_\mathrm{SOHO} = 139$ ${\mathrm{km~s^{-1}}}$, and that 
the propagation speed increases to $V_\mathrm{bg} = 327$ ${\mathrm{km~s^{-1}}}$ 
with radial distance. This ICME shows acceleration, and the fit has a value of ${\alpha} = 0.276$. 

      %
      %
      %
     
      %
      \begin{figure}
      \begin{center}
      \centerline{\includegraphics[width=0.8\textwidth,clip=]{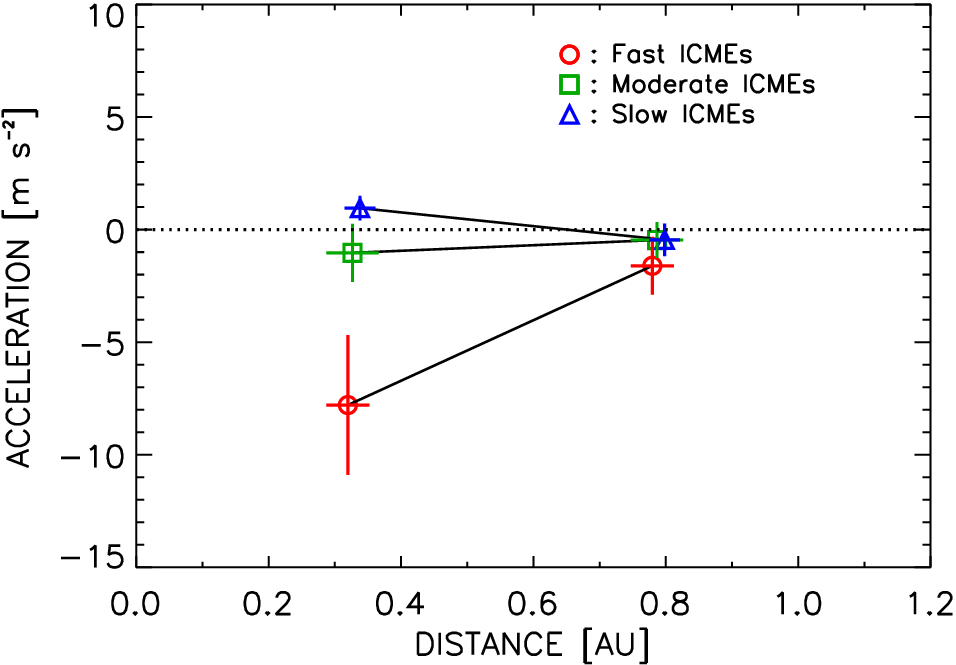}}
      \vspace{-0.05 \textwidth}
      \caption{
      Average radial evolution of acceleration for the fast ($V_{\mathrm{SOHO}} - V_\mathrm{bg} > 500$ ${\mathrm{km~s^{-1}}}$), moderate 
      ($0$ ${\mathrm{km~s^{-1}}}$ $\le V_{\mathrm{SOHO}} - V_\mathrm{bg} \le 500$ ${\mathrm{km~s^{-1}}}$), and slow 
      ($V_{\mathrm{SOHO}} - V_\mathrm{bg} < 0$ ${\mathrm{km~s^{-1}}}$) ICMEs in this study. Average accelerations are derived from 
      Equations~(\ref{eq.accel1}) and (\ref{eq.accel2}) with reference distances [${R_\mathrm{1}}$ and ${R_\mathrm{2}}$] for each 
      ICME. Open circle, square, and triangle symbols indicate data points that consist of [${R_\mathrm{1}}$, ${a_\mathrm{1}}$] and 
      [${R_\mathrm{2}}$, ${a_\mathrm{2}}$] averaged for 14 fast, 20 moderate, and 5 slow ICMEs, respectively. 
      Pairs of symbols are connected by solid lines.
      }
      \label{fig6}
      \end{center}
      \end{figure}
     
Figure \ref{fig6} shows the average radial acceleration for groups of fast, moderate, and slow 
ICMEs; the average acceleration in the two regions [$a_\mathrm{1}$ and $a_\mathrm{2}$] are calculated first using 
Equations (\ref{eq.accel1}) and (\ref{eq.accel2}) for each ICME, and each is subsequently averaged for respective 
groups. For all of them, the mean values of ${R_\mathrm{1}}$ and ${R_\mathrm{2}}$ with the standard errors are 
$0.33 \pm 0.04$ and $0.79 \pm 0.04$ AU, respectively. From this figure, we confirm that the acceleration 
levels vary toward zero with an increase in distance, and this trend is conspicuous for the group of fast ICMEs. 
We also confirm that the group of moderate ICMEs shows little acceleration.

\subsection{Critical Speed for Zero Acceleration}
      \label{criticalspeed}

\begin{figure}
      \begin{center}
      \centerline{\includegraphics[width=0.8\textwidth,clip=]{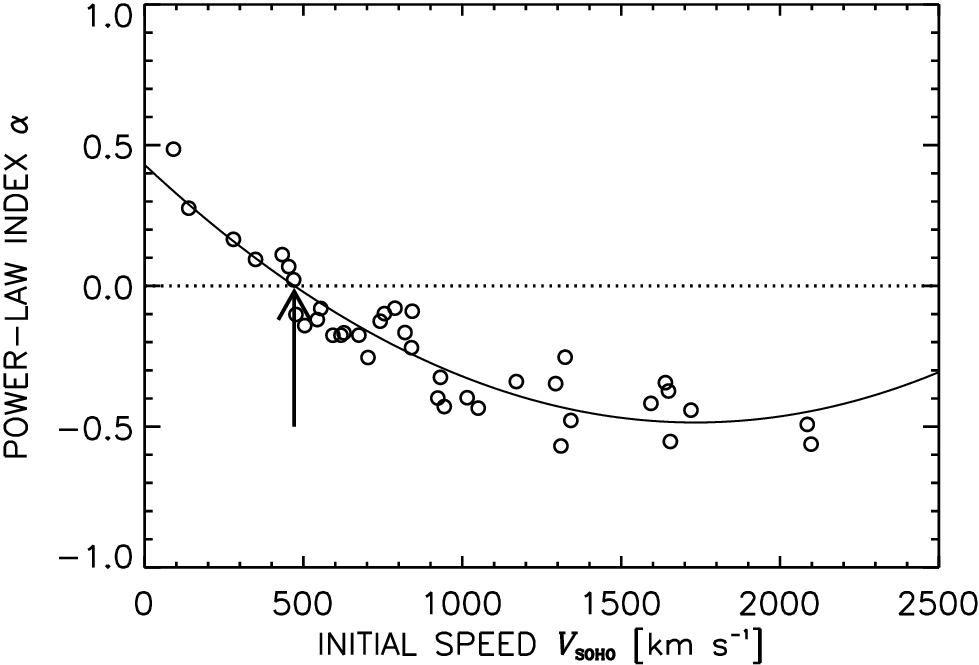}}
      \vspace{-0.05 \textwidth}
      \caption{
      Relationship between estimated initial speeds [${V_\mathrm{SOHO}}$] and indices [${\alpha}$] for Equation (\ref{eq.powerlaw}) 
      for the 39 ICMEs in this study. Solid and dotted lines show the best-fit quadratic curve 
      ${\alpha} = k_\mathrm{1} + k_\mathrm{2}V_\mathrm{SOHO} + k_\mathrm{3}V_{\mathrm{SOHO}}^{\mathrm{2}}$ and the ${\alpha} = 0$ line. 
      The intersection point of these lines is indicated by an arrow, and corresponds to the critical speed for zero acceleration 
      [${V_\mathrm{c1}}$], which is $471 \pm 19$ ${\mathrm{km~s^{-1}}}$.
      }
      \label{fig7}
      \end{center}
      \end{figure}

\begin{figure}
      \begin{center}
      \centerline{\includegraphics[width=0.8\textwidth,clip=]{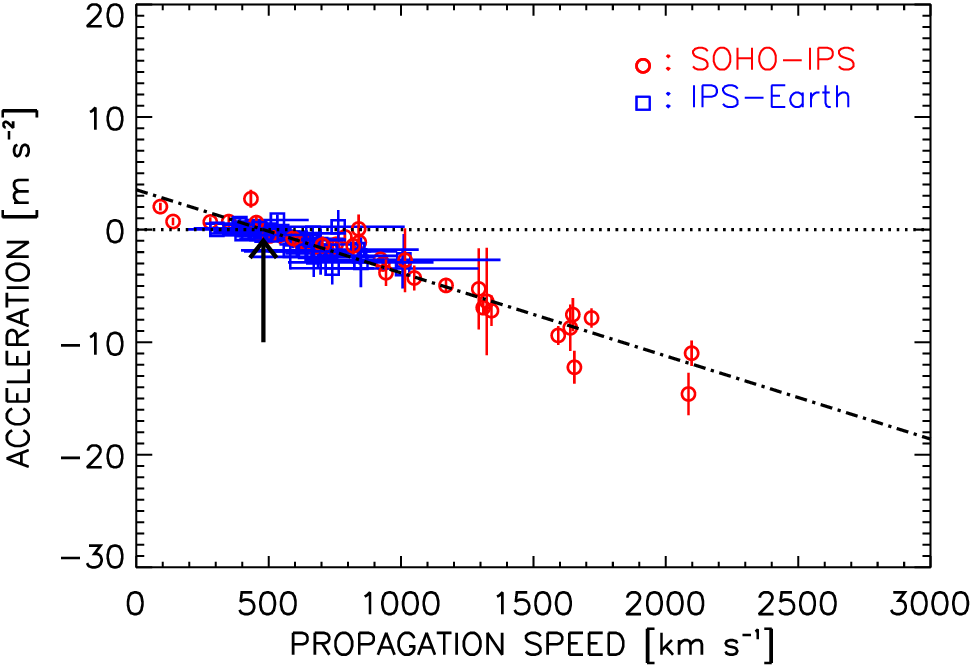}}
      \vspace{-0.05 \textwidth}
      \caption{
      Relationship between propagation speeds and accelerations for the 39 ICMEs in this study. Accelerations 
      are derived from Equations (\ref{eq.accel1}) and (\ref{eq.accel2}), while values of ${V_\mathrm{SOHO}}$ and 
      ${V_\mathrm{IPS}}$ are used for the propagation speeds. Open circle and square symbols denote data points, 
      which are [${V_\mathrm{SOHO}}$, ${a_\mathrm{1}}$] for the SOHO--IPS region and [${V_\mathrm{IPS}}$, ${a_\mathrm{2}}$] 
      for the IPS--Earth region, respectively. Dash--dotted and dotted lines show the best-fit line and 
      zero acceleration line, respectively. The arrow indicates the critical speed for zero acceleration [${V_\mathrm{c2}}$], which 
      is $480 \pm 21$ ${\mathrm{km~s^{-1}}}$.
      }
      \label{fig8}
      \end{center}
      \end{figure}

If ICMEs accelerate or decelerate by interaction with the solar wind, we expect that the 
acceleration will become zero when the propagation speed of ICMEs reaches the speed 
of the background solar wind. Therefore, it is important to know the ICME propagation 
speed in this situation in order to verify our expectations. Here, we call this speed 
``the critical speed for zero acceleration''. In Figures \ref{fig7} and \ref{fig8}, 
we give information on this critical speed for zero acceleration in two ways. 
In Figure \ref{fig7}, we show the relationship between initial ICME speeds [${V_\mathrm{SOHO}}$] and 
indices [${\alpha}$]. The ${\alpha}$ indicates the type of ICME motion, \textit{i.e.} 
acceleration (${\alpha} > 0$), uniform (${\alpha} = 0$), and deceleration (${\alpha} < 0$). 
As shown here, ${\alpha}$ ranges from $0.486$ to $-0.596$ with an increase in ${V_\mathrm{SOHO}}$. 
Table \ref{table3} gives the mean values of the critical speed for zero acceleration [${V_\mathrm{c1}}$], 
coefficients [${k_\mathrm{1}}$, ${k_\mathrm{2}}$, and ${k_\mathrm{3}}$] for the best-fit curve, 
and their standard errors. Figure \ref{fig8} shows the relationship between ICME speeds [${V_\mathrm{SOHO}}$ 
and ${V_\mathrm{IPS}}$] and accelerations [${a_\mathrm{1}}$ and ${a_\mathrm{2}}$]. 
Table \ref{table4} presents the mean values of the critical speed for zero acceleration [${V_\mathrm{c2}}$] 
slope, and intercept for the best-fit line and their standard errors, which are estimated 
using the \textsf{FITEXY.pro} from the IDL Astronomy User's Library (\url{idlastro.gsfc.nasa.gov/homepage.html}). 
From the above examinations, we find $V_\mathrm{c1} = 471 \pm 19$  ${\mathrm{km~s^{-1}}}$ and 
$V_\mathrm{c2} = 480 \pm 21$ ${\mathrm{km~s^{-1}}}$ as the critical speed for zero acceleration.

      %
      %

     %
      \begin{table}
      \caption{
      Mean values of coefficients [${k_\mathrm{1}}$, ${k_\mathrm{2}}$, and ${k_\mathrm{3}}$] for the best-fit 
      quadratic curve ${\alpha} = k_\mathrm{1} + k_\mathrm{2}V_\mathrm{SOHO} + k_\mathrm{3}V_{\mathrm{SOHO}}^{2}$ 
      and the critical speed for zero acceleration [${V_\mathrm{c1}}$], and their standard errors, which were derived 
      from the relationship between ${V_\mathrm{SOHO}}$ and ${\alpha}$.
      }
     \label{table3}
     \begin{tabular}{cccccc}
      \hline
~~~ & ${k_\mathrm{1}}$ & ${k_\mathrm{2}}$ & ${k_\mathrm{3}}$ & ${V_\mathrm{c1}}$~[${\mathrm{km~s^{-1}}}$] \\
      \hline
Mean & ${4.31 \times 10^{-1}}$ & ${-1.06 \times 10^{-3}}$ & ${3.04 \times 10^{-7}}$ & 471 \\
Standard error & ${5.58 \times 10^{-2}}$ & ${1.16 \times 10^{-4}}$ & ${5.22 \times 10^{-8}}$ & 19 \\
      \hline
      \end{tabular}
      \end{table}
     %
      \begin{table}
      \caption{
     Mean values of slope and intercept for the best-fit line and the critical speed for zero acceleration 
     [${V_\mathrm{c2}}$] and their standard errors, which were derived from the relationship between speeds and 
     accelerations of ICMEs.
      }
     \label{table4}
     \begin{tabular}{ccccc}
      \hline
~~~ & Slope [${\mathrm{s^{-1}}}$] & Intercert [${\mathrm{m~s^{-2}}}$] & ${V_\mathrm{c2}}$~[${\mathrm{km~s^{-1}}}$] \\
      \hline
Mean & ${-7.38 \times 10^{-6}}$ & 3.54 & 480 \\
Standard error & ${2.03 \times 10^{-7}}$ & ${1.24 \times 10^{-1}}$ & 21 \\
      \hline
      \end{tabular}
      \end{table}

\subsection{Relationship Between Acceleration and Difference in Speed}
      \label{speeddifference}
    
\begin{figure}
      \centerline{\includegraphics[width=0.9\textwidth,clip=]{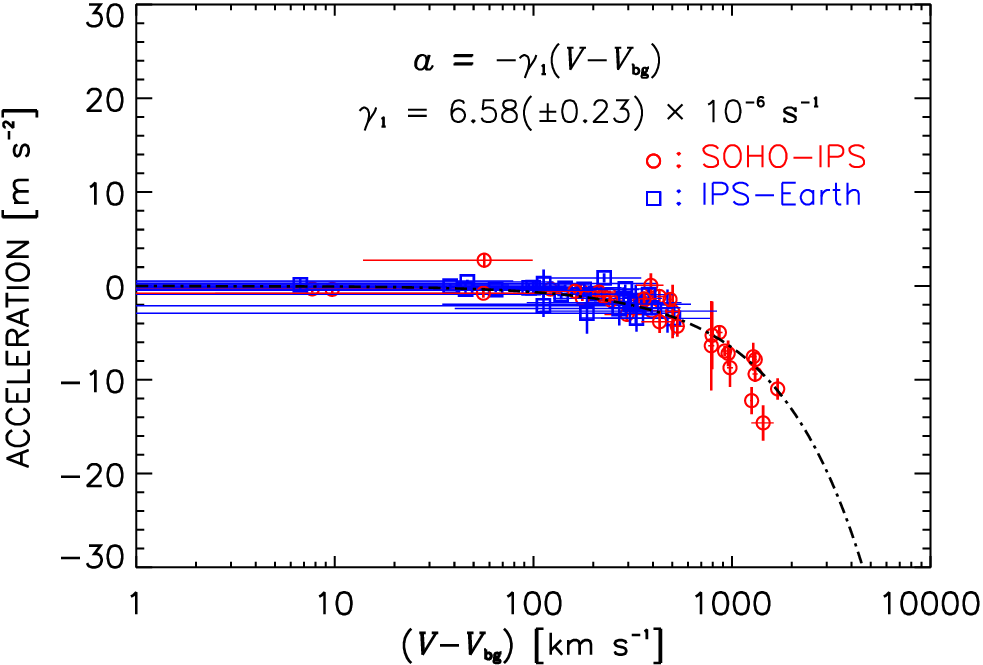}}
      \vspace{-0.05\textwidth}
      \leftline{\small \bf     
              \hspace{0.0 \textwidth}{(a)}
              \hfill
              }
      \vspace{0.05\textwidth}
      \centerline{\includegraphics[width=0.9\textwidth,clip=]{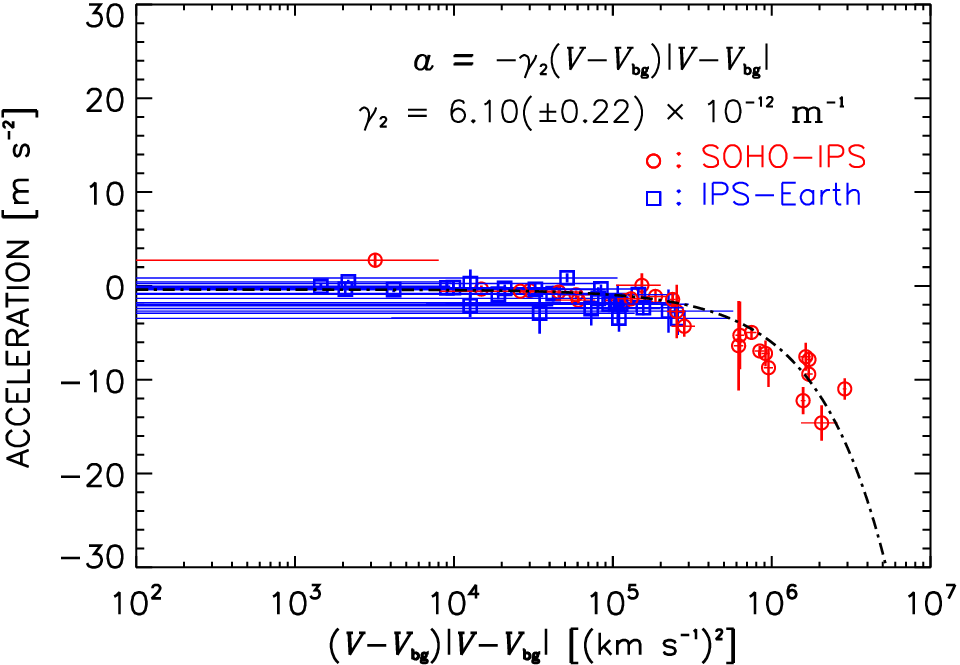}}
      \vspace{-0.05 \textwidth}
      \leftline{\small \bf     
              \vspace{0.0 \textwidth}{(b)}
              \hfill
              }
      \vspace{0.04\textwidth}
      \caption{
      Relationships between (a) acceleration [$a$] and speed difference [${V - V_\mathrm{bg}}$], and 
      (b) between $a$ and ${(V - V_\mathrm{bg})|V - V_\mathrm{bg}|}$, for 34 of the fast and moderate ICMEs 
      (\textit{i.e.} $V_{\mathrm{SOHO}} - V_\mathrm{bg} \ge 0$ ${\mathrm{km~s^{-1}}}$) in this study. 
      Open circle and square symbols denote data points that consist of values of ($V_{\mathrm{SOHO}} - V_\mathrm{bg}$) 
      and ${a_\mathrm{1}}$ [or $(V_{\mathrm{SOHO}} - V_\mathrm{bg})|V_{\mathrm{SOHO}} - V_\mathrm{bg}|$ and ${a_\mathrm{1}}$] 
      for the SOHO--IPS region and those in which consist of values of 
      ($V_{\mathrm{IPS}} - V_\mathrm{bg}$) and ${a_\mathrm{2}}$ 
      [or $(V_{\mathrm{IPS}} - V_\mathrm{bg})|V_{\mathrm{IPS}} - V_\mathrm{bg}|$ and ${a_\mathrm{2}}$] for 
      the IPS--Earth region, respectively. 
      In each panel, the dash--dotted curve denotes the best-fit line shown as a curve because of the logarithmic $x$-axis
      scale.
      }
     \label{fig9}
     \end{figure}

We investigated how the ICME acceleration relates to the difference in speed between 
it and the background solar wind. In this investigation, we attempted to show which is 
more suitable to describe the relationship between acceleration and difference in speed: 
$a = -{\gamma}_{\mathrm{1}}(V - V_\mathrm{bg})$ or $a = -{\gamma}_{\mathrm{2}}(V - V_\mathrm{bg})|V - V_\mathrm{bg}|$; 
these expressions were introduced and also tested in the earlier study by \inlinecite{Vrsnak2002}. 
Here, $a$, $V$, and ${V_\mathrm{bg}}$ denote the acceleration, ICME speed, and speed of 
the background solar wind, respectively. Although it was assumed that the coefficients 
[${\gamma}_{\mathrm{1}}$ and ${\gamma}_{\mathrm{2}}$] decrease with the heliocentric 
distance in the earlier study, for this analysis we assume that the values of coefficients are 
constants because we want as few variables as possible to describe the relationship. 
We also assume that the speed of the background solar wind [${V_\mathrm{bg}}$] is constant for heliocentric 
distances ranging from ${\approx}~0.1$ to 1 AU. This assumption has been verified approximately between 
0.3 and 1 AU by \inlinecite{Neugebauer1975} and \inlinecite{Schwenn1981}. 
In Figure \ref{fig9}, the top panel shows the relationship between $a$ and $(V - V_\mathrm{bg})$, and the bottom 
panel that between $a$ and $(V - V_\mathrm{bg})|V - V_\mathrm{bg}|$ for ICMEs with 
$(V_\mathrm{SOHO} - V_\mathrm{bg}) \ge 0$ ${\mathrm{km~s^{-1}}}$, 
\textit{i.e.} the fast and moderate ICMEs. Table \ref{table5} exhibits the values of ${{\gamma}_{\mathrm{1}}}$ 
and ${{\gamma}_{\mathrm{2}}}$, correlation coefficients, and reduced ${\chi^{\mathrm{2}}}$ derived from 
this analysis. It is noted that the ${{\gamma}_{\mathrm{1}}}$, ${{\gamma}_{\mathrm{2}}}$, and ${\chi^{\mathrm{2}}}$ are 
calculated using the \textsf{FITEXY.pro}. Although we also examined the slow ICMEs in the same 
way, we did not obtain a conclusive result. We discuss interpretations of these results in the 
next section.
      
     %
     
     %
      \begin{table}
      \caption{
      Coefficients [${\gamma_\mathrm{1}}$ and ${\gamma_\mathrm{2}}$], correlation coefficient [CC], 
      and reduced $\chi^{2}$ for the linear and quadratic equations.
      }
      \label{table5}
      \begin{tabular}{cccccc}
      \hline
Equation & Mean & Standard error & CC & $\chi^{2}$ \\
      \hline
\multicolumn{2}{l}{${\gamma_\mathrm{1}}$ [${\mathrm{s^{-1}}}$]} & ~~~ \\
Linear & ${6.58 \times 10^{-6}}$ & ${2.34 \times 10^{-7}}$ & ${-0.93}$ & 1.26 \\
\multicolumn{2}{l}{${\gamma_\mathrm{2}}$ [${\mathrm{m^{-1}}}$]} & ~~~ \\
Quadratic & ${6.10 \times 10^{-12}}$ & ${2.25 \times 10^{-13}}$ & ${-0.90}$ & 2.90 \\
      \hline
      \end{tabular}
      \end{table}
     

\section{Discussion} 
      \label{discussion} 

From Figures \ref{fig2}, \ref{fig3}, and \ref{fig4}, we confirm that fast and moderate ICMEs are 
rapidly and gradually decelerating during their outward propagation, respectively, while slow ICMEs are 
accelerating, and consequently all attain speeds close to those of the background solar wind. 
As shown in Figure \ref{fig5}, the distribution of ICME propagation speeds in the near-Sun region 
is wider than in the near-Earth region for all of the ICMEs identified in this study. This is 
consistent with the earlier study by \inlinecite{Lindsay1999}. We also confirm that the distribution 
of ICME propagation speed in the near-Earth region is similar to that of the background 
solar-wind speed at 1 AU. We interpret these results as indicating that ICMEs accelerate or 
decelerate by interaction with the solar wind; the magnitude of the propelling or retarding 
force acting upon ICMEs depends on the difference between ICMEs and the solar wind. 
Thus, ICMEs attain final speeds close to the solar-wind speed as they move outward from 
the Sun. Figure \ref{fig5} also shows the radial evolution of ICME propagation speeds between 0.08 
and 1 AU. We show that ICME speeds reach their final value at $0.79 \pm 0.04$ AU or at a solar 
distance slightly less than 1 AU. In addition, we confirm from Figure \ref{fig6} that the acceleration 
at $0.79 \pm 0.04$ AU is much lower than at $0.33 \pm 0.04$ AU; this is the clearest for the group of 
fast ICMEs. From this, we thus conclude that most of the ICME acceleration or deceleration 
ends by $0.79 \pm 0.04$ AU. This is consistent with an earlier result obtained by \inlinecite{Gopalswamy2001}. 

We expect that the critical speed of zero acceleration will be close to that of the background 
solar-wind speed on the basis of the above. We derive two different critical speeds of 
$V_\mathrm{c1} = 471 \pm 19$ ${\mathrm{km~s^{-1}}}$ and of $V_\mathrm{c2} = 480 \pm 21$ ${\mathrm{km~s^{-1}}}$ 
from the observational data. Although there is agreement between them, both are somewhat higher than 
the ${\approx}~380$ ${\mathrm{km~s^{-1}}}$ reported to be the threshold speed by \inlinecite{Manoharan2006} 
and the 405 ${\mathrm{km~s^{-1}}}$ reported by \inlinecite{Gopalswamy2000}. We suggest that 
this discrepancy is caused by the difference in our analysis methods and also the time 
interval chosen for the analysis. Because the properties of the background solar wind 
(\textit{e.g.} speed and density) vary with the change in solar activity, we 
consider this discrepancy to be minor, and we note that both critical speeds in our result are 
within the typical speed of the solar wind: $V_\mathrm{bg} = 445 \pm 95$ ${\mathrm{km~s^{-1}}}$ 
from our sample. Here, we adopt the speed of 480 ${\mathrm{km~s^{-1}}}$ as the critical speed 
for zero acceleration as a mean that is derived from the relationship between 
propagation speeds and accelerations without the assumption of a power-law form 
for the motion of the ICME. 

\inlinecite{Vlasov1992} and \inlinecite{Manoharan2006} point out that the radial evolution of ICME speeds 
can be represented by a power-law function. A power-law speed evolution also applies to 
the ICMEs identified in this study as shown in Figures \ref{fig2}, \ref{fig3}, and \ref{fig4}. 
As indicated by Figure \ref{fig7}, the value of ${\alpha}$ varies from 0.499 (acceleration) to 
${-0.596}$ (strongly deceleration) as ICME speeds increase. 
This result is consistent with that exhibited in Figure \ref{fig8}. 

The relationship between acceleration and speed-difference for ICMEs is usually expressed 
by either of the following: a linear equation $a = -{\gamma}_{\mathrm{1}}(V - V_\mathrm{bg})$ 
or a quadratic equation $a = -{\gamma}_{\mathrm{2}}(V - V_\mathrm{bg})|V - V_\mathrm{bg}|$. 
As shown in Figure \ref{fig9}, these equations are evaluated using 
the acceleration and speed-difference data derived from our observations. From this and 
Table \ref{table5}, we find that the reduced ${\chi^{2}}$ for the former relationship is 
smaller than for the latter. The assessment of the significance level shows that ${\chi}^{2} = 1.26$ 
for the linear equation is smaller than the reduced ${\chi^{2}}$ corresponding to 
the probability of 0.05 with 66 degrees of freedom, while ${\chi}^{2} = 2.90$ for the quadratic 
one is larger. We therefore conclude that the linear equation is more suitable than 
the quadratic one to describe the kinematics of ICMEs with 
$(V_\mathrm{SOHO} - V_\mathrm{bg}) \ge 0$ ${\mathrm{km~s^{-1}}}$. 
From the viewpoint of fluid dynamics, a linear equation suggests that the hydrodynamic 
Stokes drag force is operating, while the quadratic equation suggests 
the aerodynamic drag force. \inlinecite{Maloney2010} found that the acceleration of a 
fast ICME showed a linear dependence on the speed difference, while that of a slow ICME 
showed a quadratic dependence. Our conclusion is consistent with their finding only for the 
fast and moderate ICMEs. We could not verify their result for the slow ICMEs because we 
lack sufficient observational data for the slow ICMEs in our sample. We expect to make a 
more detailed examination for the motion of slow ICMEs in a future study. 

We also obtained the mean value of $6.58 \times 10^{-6}~\mathrm{s^{-1}}$ for the coefficient 
${{\gamma}_{\mathrm{1}}}$ in our analysis. Substituting our value of ${{\gamma}_{\mathrm{1}}}$ 
in our linear equation, we obtain the following simple expression: 
\begin{equation}
   \label{eq.kinematics}
a = -6.58 \times 10^{-6}(V - V_\mathrm{bg}),
\end{equation}
where $a$, $V$, and ${V_\mathrm{bg}}$ are the acceleration, ICME propagating speed, and 
speed of the background solar wind, respectively, as a useful way to determine the dynamics of ICMEs. 

Last, we discuss why the linear equation with a constant ${{\gamma}_{\mathrm{1}}}$ can explain the observational 
result. Our IPS radio-telescope system observes fluctuations of radio signals. These fluctuations 
are proportional to the solar-wind (electron) density [${N_\mathrm{e}}$]. Therefore, low-density 
ICMEs may not be detected by our system. Moreover, we used a threshold \textit{g}-value more 
severe than that used by \inlinecite{Manoharan2006} or \inlinecite{Gapper1982} for identification of 
ICMEs. Hence, it is conceivable that almost all detected ICMEs are high-density events in 
this study. In addition, from a theoretical study, \inlinecite{Cargill2004} indicated that with dense 
ICMEs, the factor ${\gamma}$ and ${C_\mathrm{D}}$ (the dimensionless drag coefficient) become approximately 
constant for aerodynamic drag deceleration; here, ${\gamma}C_\mathrm{D} = {\gamma}_{\mathrm{2}}$ 
in our notation. From this, we surmise that a constant value of ${{\gamma}C_\mathrm{D}}$ indicates that both 
interplanetary-space conditions and the properties of dense ICMEs are unchanged in the range 
from the Sun to the Earth. Therefore, ${{\gamma}_{\mathrm{1}}}$ must also become approximately 
constant over the same range from the Sun to the Earth. Thus, to recapitulate, the events detected using 
our IPS radio-telescope system give results for dense ICMEs, and the dynamics of these are well explained 
by a linear equation with ${\gamma}_{\mathrm{1}} =$ constant.

\section{Summary and Conclusions} 
      \label{conclusion} 

We investigate radial evolution of propagation speed for 39 ICMEs detected by SOHO/LASCO, 
IPS at 327 MHz, and \textit{in-situ} observations during 1997\,--\,2009 covering nearly all 
of Solar Cycle 23. In this study, we first analyze \textit{g}-values obtained by STEL IPS observations 
in the above period, and find 497 IPS disturbance event days (IDEDs) as candidates 
for ICME events. Next, we compare the list of these IDEDs with that of CME/ICME pairs 
observed by SOHO/LASCO and \textit{in-situ} observations, and finally we are left with 50 ICMEs; 
those ICMEs that traveled from the Sun to the Earth, and were detected at three locations between 
the Sun and the Earth's orbit, \textit{i.e.} near-Sun, interplanetary space, and near-Earth. For 
these ICMEs, we determine reference distances and derive the propagation speeds and accelerations 
in the SOHO--IPS and IPS--Earth regions. Our examinations yield the following 
results. 

\begin{enumerate}
\item Fast ICMEs (with $V_\mathrm{SOHO} - V_\mathrm{bg} > 500$ ${\mathrm{km~s^{-1}}}$) rapidly decelerate, 
moderate ICMEs (with $0$ ${\mathrm{km~s^{-1}}}$ $\le V_\mathrm{SOHO} - V_\mathrm{bg} \le 500$ ${\mathrm{km~s^{-1}}}$) 
show either gradually deceleration or uniform motion, while slow ICMEs 
(with $V_\mathrm{SOHO} - V_\mathrm{bg} < 0$ ${\mathrm{km~s^{-1}}}$) accelerate, where ${V_\mathrm{SOHO}}$ 
and ${V_\mathrm{bg}}$ are the initial speed of ICME and the speed of the background solar wind, 
respectively. Consequently, radial speeds converge to the speed of the background solar 
wind during their outward propagation. Thus, the distribution of ICME propagation 
speeds in the near-Earth region is narrower than in the near-Sun region, as shown in 
Figure \ref{fig5}. This is consistent with the earlier study by \inlinecite{Lindsay1999}. 
\item Both the ICME accelerations and the decelerations are nearly complete by $0.79 \pm 0.04$ AU. 
This is consistent with an earlier result obtained by \inlinecite{Gopalswamy2001}. 
Both critical speeds (where the speed of ICME acceleration becomes zero) derived from 
our analysis, \textit{i.e.} $471 \pm 19$ ${\mathrm{km~s^{-1}}}$ and $480 \pm 21$ ${\mathrm{km~s^{-1}}}$, 
are somewhat higher than the values reported by \inlinecite{Manoharan2006} and \inlinecite{Gopalswamy2000}. 
However, this discrepancy is most likely explained because our analysis methods and data collection 
periods are different. Both critical speeds in our result do not differ much from the typical 
speed of the solar wind, and we adopt the mean value of 480 ${\mathrm{km~s^{-1}}}$ as the critical 
speed for zero acceleration. This is close to the speed of the background solar wind, 
$V_\mathrm{bg} = 445 \pm 95$ ${\mathrm{km~s^{-1}}}$, during this period of study. 
\item For ICMEs with $(V_\mathrm{SOHO} - V_\mathrm{bg}) \ge 0$ ${\mathrm{km~s^{-1}}}$, 
a linear equation $a = -{\gamma}_{\mathrm{1}}(V - V_\mathrm{bg})$ 
with ${\gamma}_{\mathrm{1}} = 6.58 \pm 0.23 \times 10^{-6}$ ${\mathrm{s^{-1}}}$ is more appropriate than a quadratic equation
$a = -{\gamma}_{\mathrm{2}}(V - V_\mathrm{bg})|V - V_\mathrm{bg}|$ to describe their kinematics, 
where ${\gamma}_{\mathrm{1}}$ and ${\gamma}_{\mathrm{2}}$ are coefficients, $a$, $V$, and ${V_\mathrm{bg}}$ 
are the acceleration and propagation speed of ICMEs, and the speed of the background solar wind, respectively, 
because the reduced ${\chi^{2}}$ for the linear equation satisfies the statistical significance level at 0.05, 
while the quadratic one does not.
\end{enumerate}

These results support our assumption that ICMEs are accelerated or decelerated by a drag 
force caused by an interaction with the solar wind; the magnitude of the drag force acting 
upon ICMEs depends on the difference in speed, and, thus, ICMEs attain final speeds close 
to the solar-wind speed when the force becomes zero. In particular, our result iii) suggests 
that ICMEs propagating faster than the background solar wind are controlled mainly by 
the hydrodynamic Stokes drag force. Moreover, our result iii) confirms the finding by \inlinecite{Maloney2010} 
only for the fast and moderate ICMEs that we measure. From the characteristics of the IPS observations and 
the result of \inlinecite{Cargill2004}, we conclude that the ICMEs detected by the IPS observations 
in this study are probably high-density events. A combination of the space-borne coronagraph, 
ground-based IPS, and satellite \textit{in-situ} observations serves to detect many ICMEs between 
the Sun and the Earth, and is a useful means to study their kinematics. 

%

%
\begin{acks}
The IPS observations were carried out under the solar-wind program of 
the Solar-Terrestrial Environment Laboratory (STEL) of Nagoya University. We acknowledge use of the 
SOHO/LASCO CME catalog; this CME catalog is generated and maintained at the CDAW Data Center 
by NASA and the Catholic University of America in cooperation with the Naval Research Laboratory. SOHO 
is a project of international cooperation between ESA and NASA. We thank NASA/GSFC's Space Physics 
Data Facility for use of the OMNIWeb service and OMNI data. We thank the IDL Astronomy User's Library 
for the use of IDL software. We acknowledge use of the comprehensive ICME catalog compiled by I.G. 
Richardson and H.V. Cane. We also thank B.V. Jackson for useful help and comments. 
\end{acks}

%
%
 \bibliographystyle{spr-mp-sola}
 \bibliography{fasticme}  

\end{article} 
\end{document}